\documentclass{ws-procs975x65}

\newcommand{\be}{\begin{eqnarray}}
\newcommand{\ee}{\end{eqnarray}}

\begin{document}

\title{Status of dark matter in the universe
\footnote{Based  on a presentation  at  the  Fourteenth  Marcel  Grossmann  Meeting  on  General  Relativity,
Rome, July 2015.}
}

\author{Katherine Freese}
\address{Physics Department, University of Michigan,
Ann Arbor, MI 48109, USA
and\\
Oskar Klein Centre for Cosmoparticle Physics, Stockholm University, Stockholm, Sweden}


\begin{abstract}
Over the past few decades, a consensus picture has emerged
in which roughly a quarter of the universe consists of
dark matter. I begin with a review of the observational evidence for the existence of
dark matter: rotation curves of galaxies, gravitational
lensing measurements, hot gas in clusters, galaxy formation,
primordial nucleosynthesis and cosmic microwave background observations.
Then I discuss a number of anomalous signals in a variety of data sets that may
point to discovery, though all of them are controversial.  The
annual modulation in the DAMA detector and/or  the gamma-ray excess
seen in the Fermi Gamma Ray Space Telescope from the Galactic Center could be due to WIMPs;
a 3.5 keV X-ray line from  multiple sources could be due to sterile neutrinos;
or the 511 keV line in INTEGRAL data could be due to MeV dark matter.
All of these would require further confirmation in
other experiments or data sets to be proven correct.
In addition, a new line of research on dark stars is presented,
which suggests that the first stars to exist in the universe
were powered by dark matter heating rather than by fusion:
the observational possibility of discovering dark matter in this
way is discussed.
\end{abstract}


\bodymatter

\section{Introduction}
A standard model of cosmology is emerging (often dubbed the
Concordance Model), in which the universe consists of 5\% ordinary
baryonic matter, $\sim 26$\% dark matter, and $\sim 69$\% dark energy.
\cite{Komatsu:2008hk,concordo2}  The baryonic content is
well-known, both from element abundances produced in primordial
nucleosynthesis roughly 100 seconds after the Big Bang, and from
measurements of anisotropies in the cosmic microwave background (CMB).
The evidence for the existence of dark matter is overwhelming, and
comes from a wide variety of astrophysical measurements.

\section{Dark Matter in Galaxies and Clusters}

\subsection{The Beginnings of the Dark Matter Problem and Rotation Curves}

The dark matter problem is perhaps the longest outstanding problem in all of modern physics.
The puzzle dates back to the 1930's, to the work first of Knut Lundmark in Sweden and shortly after that Fritz Zwicky at Caltech.
Zwicky noticed that galaxies in the Coma Cluster were moving too rapidly to be explained by the stellar
material in the cluster.  He postulated that additional mass in the form of something dark must
be providing the gravitational pull to speed up the orbits.  Subsequent work continued to find
similar evidence, but it wasn't until the work of Ford and Rubin \cite{FordRubin1970} in the 1970's 
that the same unexplained rapid orbits were found
to exist in every single galaxy.  At that point the scientific consensus for dark matter emerged.
For a review of dark matter history, see the review of Ref.~\refcite{Bertone:2016nfn}.

Rotation curves of
galaxies are flat.  The velocities of objects (stars or gas) orbiting
the centers of galaxies, rather than decreasing as a function of the
distance from the galactic centers as had been expected, remain
constant out to very large radii.  Similar observations of flat
rotation curves have now been found for all galaxies studied,
including our Milky Way.  The simplest explanation is that galaxies
contain far more mass than can be explained by the bright stellar
objects residing in galactic disks.  This mass provides the force to
speed up the orbits.  To explain the data, galaxies must have enormous
dark halos made of unknown `dark matter.' Indeed, more than 95\% of
the mass of galaxies consists of dark matter.  This is illustrated in
Fig. 1, where the velocity profile of galaxy NGC 6503 is displayed as
a function of radial distance from the galactic center. The baryonic
matter which accounts for the gas and disk cannot alone explain the
galactic rotation curve. However, adding a dark matter halo allows a
good fit to data.\footnote{It is interesting to note that alternative scenarios without dark matter
began with modified Newtonian dynamics (MOND). \cite{milgrom}  While these models
have been shown to fail, particularly by cosmic microwave background observations, 
they may provide an interesting phenomenological fit on small scales. \cite{mcgaugh}}

The limitations of rotation curves are that one can only look out as
far as there is light or neutral hydrogen (21 cm), namely to distances
of tens of kpc.  Thus one can see the beginnings of dark matter haloes, but
cannot trace where most of the dark matter is. The lensing experiments
discussed in the next section go beyond these limitations.

\begin{figure}
\includegraphics[width=\textwidth]{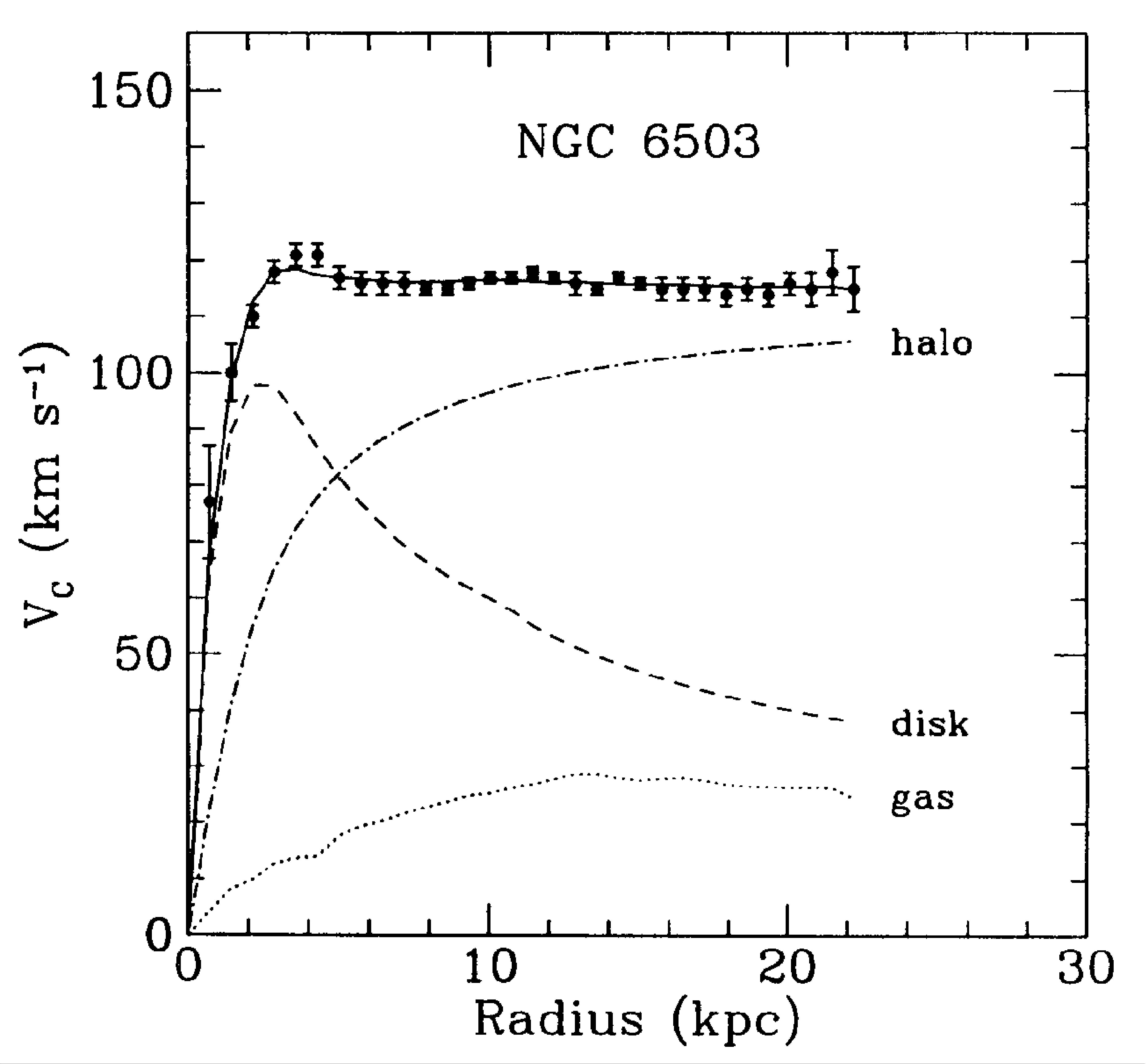}
\caption{Galactic rotation curve for NGC 6503 showing disk and gas
  contribution plus the dark matter halo contribution needed to match
  the data.}
\end{figure}

\subsection{Gravitational Lensing}

Einstein's theory of general relativity predicts that mass bends, or
lenses, light.  This effect can be used to gravitationally ascertain
the existence of mass even when it emits no light.  Lensing
measurements confirm the existence of enormous quantities of dark
matter both in galaxies and in clusters of galaxies.

Observations are made of distant bright objects such as galaxies or
quasars.  As the result of intervening matter, the light from these
distant objects is bent towards the regions of large mass.  Hence
there may be multiple images of the distant objects, or, if these
images cannot be individually resolved, the background object may
appear brighter.  Some of these images may be distorted or sheared.
The Sloan Digital Sky Survey used weak lensing (statistical studies of
lensed galaxies) to conclude that galaxies, including the Milky Way,
are even larger and more massive than previously thought, and require
even more dark matter out to great distances. \cite{Sloan2005}  Again, the
predominance of dark matter in galaxies is observed.

A beautiful example of a strong lens is shown in Fig.~2.  The panel
on the right shows a computer reconstruction of a foreground cluster
inferred by lensing observations made by Tyson et al.\ \cite{tyson} using the Hubble
Space Telescope.  This extremely rich cluster contains many galaxies,
indicated by the peaks in the figure.  In addition to these galaxies,
there is clearly a smooth component, which is the dark matter
contained in clusters in between the galaxies.

The key success of the lensing of dark matter to date is the evidence that dark matter
is seen out to much larger distances than could be probed by rotation
curves: the dark matter is seen in galaxies out to 200 kpc from the centers
of galaxies, in agreement with
$N$-body simulations.  On even larger Mpc scales, there is
evidence for dark matter in filaments (the cosmic web).

\begin{figure}
\includegraphics[width=\textwidth]{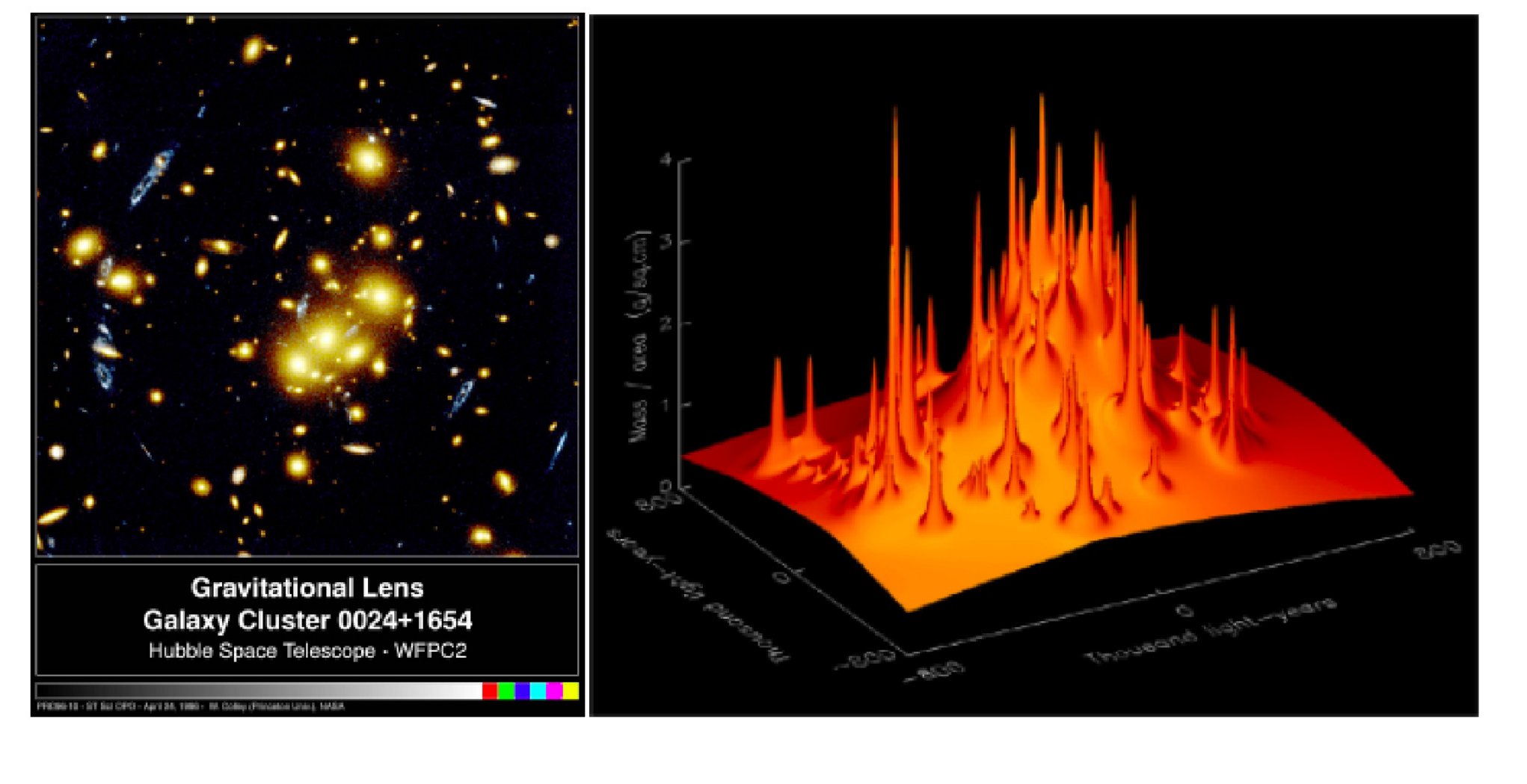}
\caption{Left: The foreground cluster of galaxies gravitationally
  lenses the blue background galaxy into multiple images. Right: A
 computer reconstruction of the lens shows a
  smooth background component not accounted for by the mass of the
  luminous objects.}
\end{figure}

\subsection{Hot Gas in Clusters}

Another piece of gravitational evidence for dark matter is the hot gas
in clusters.  Fig.~3 illustrates the Coma Cluster. The left panel is
in the optical, while the right panel is emission in the X-ray
observed by ROSAT. \cite{Coma1997}
[Note that these two images are not
on the same scale.]  The X-ray image indicates the presence of hot
gas.  The existence of this gas in the cluster can only be explained
by a large dark matter component that provides the potential well to
hold on to the gas.

\begin{figure}
\includegraphics[width=0.49\textwidth]{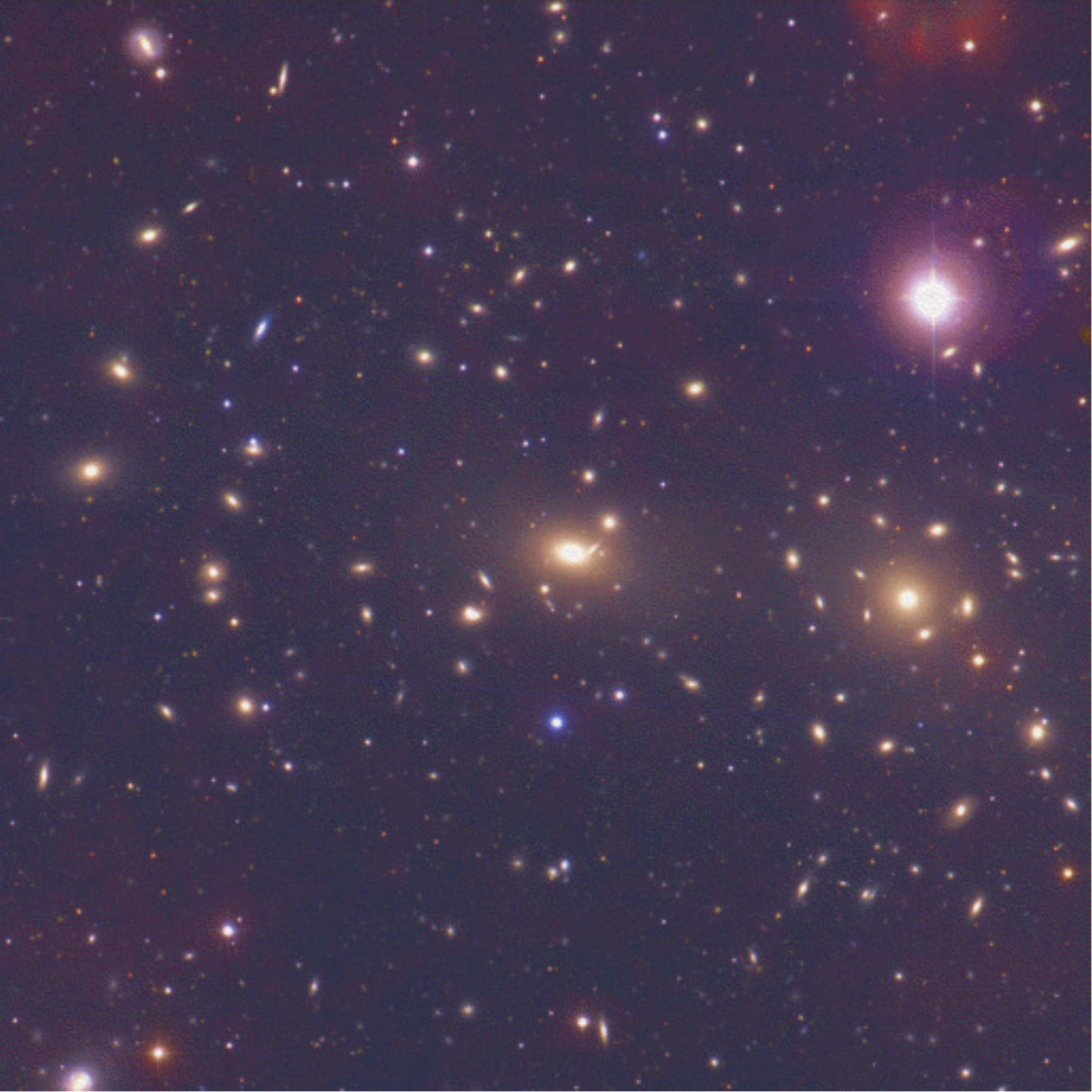}
\includegraphics[width=0.49\textwidth]{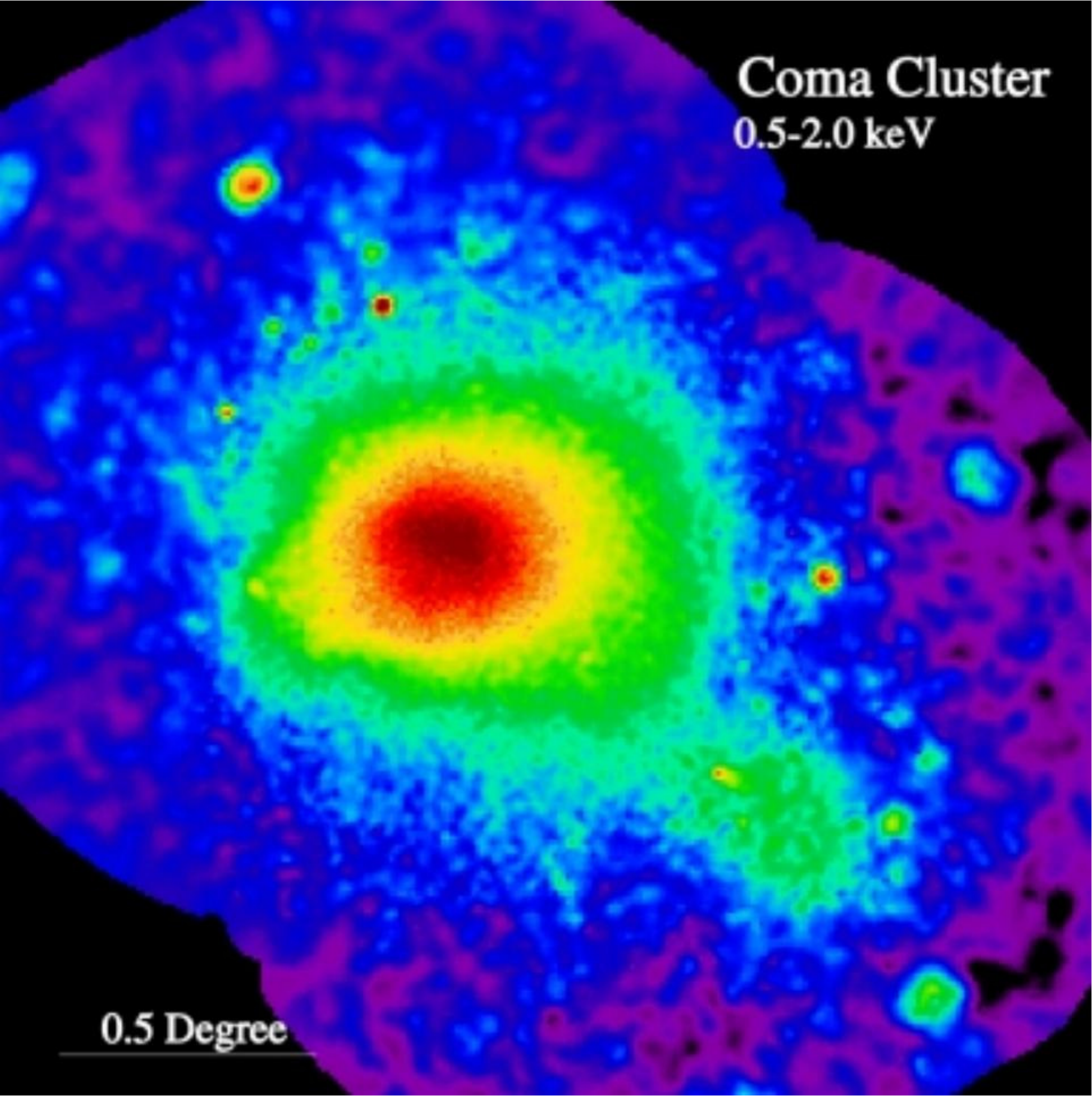}
\caption{COMA Cluster: without dark matter, the hot gas would
  evaporate. Left panel: optical image. Right panel: X-ray image from
  ROSAT satellite.}
\end{figure}

\subsection{Bullet Cluster}

An image (shown in Fig.~4) of the Bullet Cluster of galaxies (a cluster formed out
of a collision of two smaller clusters) taken by the Chandra X-ray
observatory shows in pink the baryonic matter; in blue is an image of
the dark matter, deduced from gravitational lensing. In the process of
the merging of the two smaller clusters, the dark matter has passed
through the collision point, while the baryonic matter slowed due to
friction and coalesced to a single region at the center of the new
cluster.  The Bullet Cluster provides clear evidence of the existence
of  two different types
of matter: baryons and dark matter behave differently.

\begin{figure}
\begin{center}
\includegraphics[width=7cm]{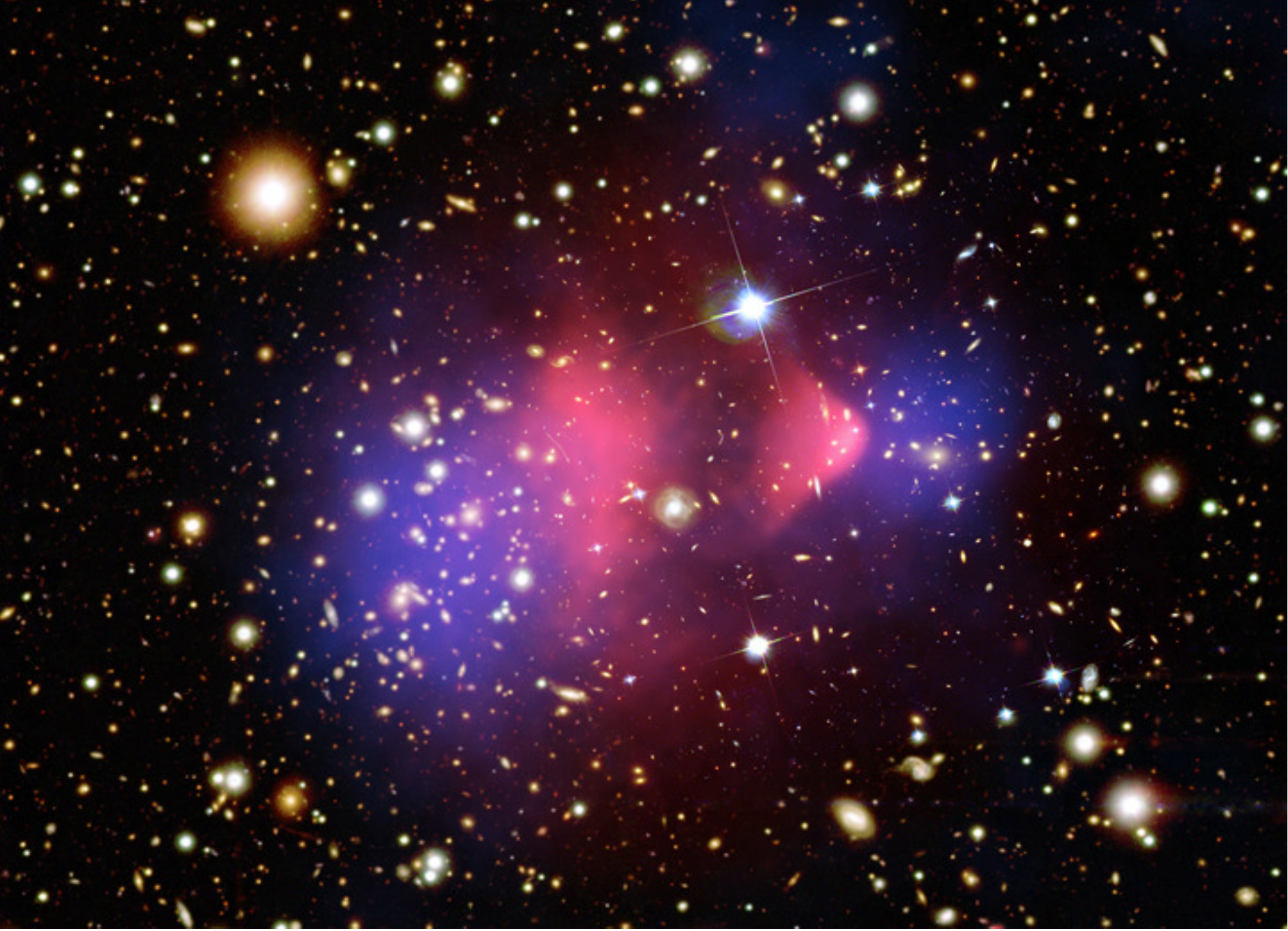}
\end{center}
\caption{The Bullet Cluster: A collision of galactic clusters shows
  baryonic matter (pink) as separate from dark matter (blue), whose
  distribution is deduced from gravitational lensing.}
\end{figure}

Thus the evidence that most of the mass of galaxies and clusters is
made of
some unknown component of dark matter is overwhelming.
As I've shown, dark matter shows its existence gravitationally in many ways, including rotation
curves (out to tens of kpc), gravitational lensing (out to 200
kpc), hot gas in clusters, and the Bullet Cluster.

Additionally,
without dark matter, large scale structure could not have formed by the present time
and we would not exist.  Until recombination at $z=1100$, the universe is ionized, baryons are
tied to photons, and both photons and baryons stream out of structures as they are forming.  It is the dark matter
that clumps together first, before recombination, and provides the potential wells for the ordinary matter to fall
into at a later time.  In order for dark matter to initiate the formation of galaxies and clusters,
it must be cold rather than hot.  Hot dark matter would be moving relativistically and would
stream out of structures in the same way that photons do; hence it was already known in the
1980s that neutrinos cannot
provide the potential wells for structure formation and cannot constitute the dark matter.
Nonrelativistic cold dark matter has become the standard paradigm for the dark matter in the universe.\footnote{Alternatives do exist including warm dark matter.}

Below I turn to the cosmic microwave background which
provides  irrefutable evidence for dark matter.

\section{Cosmic Abundances}
The cosmic abundances tell a consistent story in which the
preponderance of the mass in the universe consists of an unknown dark matter
component.  The cosmic microwave background provides the most powerful
measurements of the cosmological parameters; primordial
nucleosynthesis restricts the abundance of baryonic matter; Type IA
supernovae provided the first evidence for the acceleration of the
universe, possibly explained by dark energy as the major constituent
of the cosmic energy density.

\subsection{The Cosmic Microwave Background}

Further evidence for dark matter comes from measurements on
cosmological scales of anisotropies in the cosmic microwave background. 
\cite{Komatsu:2008hk,concordo2}  The CMB is the remnant
radiation from the hot early days of the universe. The photons
underwent oscillations that froze in just before decoupling from the
baryonic matter at a redshift of 1100.  The angular scale and height
of the peaks (and troughs) of these oscillations are powerful
probes of cosmological parameters, including the total energy density,
the baryonic fraction, and the dark matter component, as shown in Fig.~5.  The sound
horizon at last scattering provides a ruler stick for the geometry of
the universe: if the light travels in a straight line (as would be the
case for a flat geometry), then the angular scale of the first Doppler
peak was expected to be found at 1 degree; indeed this is found to be
correct.  Thus the geometry is flat, corresponding to an energy
density of the universe of $\sim 10^{-29} {\rm gm/cm}^3$.  The height
of the second peak implies that 5\% of the total is ordinary atoms,
while matching all the peaks implies that 26\% of the total is dark matter.
Indeed the CMB by itself provides irrefutable evidence for dark matter.

\begin{figure}
\begin{center}
\includegraphics[width=7cm]{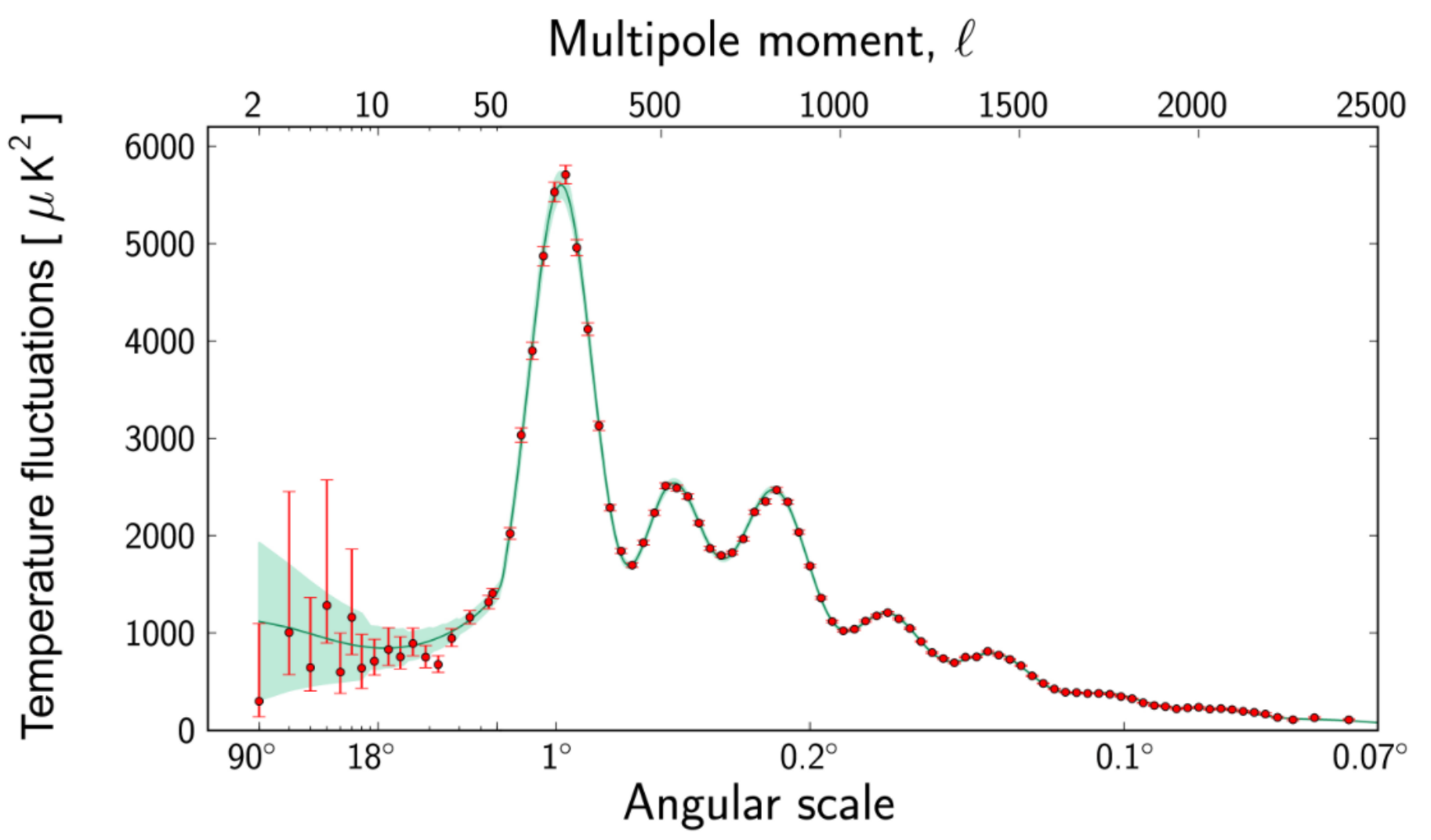}
\end{center}
\caption{Planck's power spectrum of temperature fluctuations in the cosmic microwave background.
The fluctuations are shown at different angular scales on the sky.  Red dots with error bars are the
Planck data.  The green curve represents the standard model of cosmology, $\Lambda$CDM. The
peak at 1 degree is consistent with a flat geometry of the universe, the height of the second peak with 5\%, and the second and third peaks with 26\% dark matter. }
\end{figure}

\subsection{Primordial nucleosynthesis}
When the universe was a few hundred seconds old, at a temperature of
ten billion degrees, deuterium became stable: $p + n \rightarrow D +
\gamma$.  Once deuterium forms, helium and lithium form as well. The
formation of heavier elements such as C, N, and O must wait a billion
years until stars form, with densities high enough for triple
interactions of three helium atoms into a single carbon atom. The
predictions from the Big Bang are 25\% Helium-4, $10^{-5}$
deuterium, and $10^{-10}$ Li-7 abundance by mass.  These predictions
exactly match the data as long as atoms are only 5\% of the total
constituents of the universe.

\subsection{Dark Energy}
The first evidence for the $\sim$70\% dark energy in the universe came from
observations of distant supernovae (Perlmutter et al.,
\cite{sn1999a} Riess et al., \cite{sn1999b} Riess et
  al.\ \cite{sn2004}). The supernovae are dimmer than expected, as
is most easily explained by an accelerating universe.  There are two
different theoretical approaches currently pursued  to explain the dark energy: (i) a vacuum energy such as a
cosmological constant or time-dependent vacuum \cite{fafm1987} may be responsible, or (ii) it is possible that
General Relativity is incomplete and that Einstein's equations need to
be modified. \cite{modgenrel2002a,modgenrel2005,modgenrel2002b,Carroll_etal2004}  Note, however, that
this dark energy does not resolve or contribute to the question of
dark matter in galaxies, which remains as puzzling (if not more) than
twenty years ago.  We now have a concordance model of the universe, in
which roughly a quarter of its content consists of dark matter.

\section{Dark Matter Candidates}

\subsection{MACHOs}

Twenty years ago, it seemed reasonable that dark matter might consist of faint stars, substellar objects,
or stellar remnants (white dwarfs or neutron stars),
i.e., stars that simply were too faint to have yet been discovered. These fall into the category
of massive compact halo objects, or MACHOs.  Other MACHO candidates would include
 primordial black holes or mirror matter. \cite{MohapatraTeplitz1999}

A combination of theory and observation have ruled these out
as solving the dark matter problem of the Milky Way.  First, 
Refs.~\refcite{faintstars,graff1996} used HST data to show that low mass stars could be at most 3\% of the Milky Way dark matter.  Next, a combination of theory plus Hipparchos parallax data was 
used to rule out substellar objects, or brown dwarfs, as the primary constituent of the Galaxy's dark matter. \cite{browndwarfs}  Stellar 
remnants were also potential DM candidates. Bounds on white dwarfs (WD) as dark matter came from many arguments (see Refs.~\refcite{machoreview,machoreview2} for a review).  Stellar precursors of white dwarfs would have produced too much IR radiation that would have swallowed TeV gamma-rays seen from objects like Markarian 451; a too large fraction of the Universe's baryonic mass budget would have been required to produce the progenitor stars of the white dwarfs; WD would have overproduced carbon and nitrogen.   From
 these constraints we argued that at most 15\% of the Milky Way Halo could be
made of white dwarfs (Freese et al.,
\cite{Freese_etal2000} Fields et al., \cite{ffgwpb},
Graff et al.\ \cite{ffgwpc}); at that time we disagreed with claims made by
 the MACHO microlensing experimental  that 100\% of the dark matter could be in the
 form of MACHOs (the experiments originally overestimated the MACHO contribution).

  Microlensing experiments (the MACHO (Alcock et al.\
\cite{alcock2000}) and EROS experiments (Ansari et al.\
\cite{eros2004}))  eventually showed that MACHOs less massive than 0.1 $M_\odot$
make an insignificant contribution to the energy density of the Galaxy.  However, there is a possible detection
(Alcock et al. \cite{alcock2000}) of a roughly 15\% halo
fraction made of $\sim 0.5 M_\odot$ objects which might be made of
stellar remnants such as white dwarfs.  These estimates agree with the numbers we found earlier
from a combination of theory and other data sets. \cite{machoreview,machoreview2}  
The white dwarf contribution to the dark matter halo could be significant, yet
not enough to explain all of the dark matter of the Milky Way.

\subsection{Nonbaryonic Dark Matter}
From primordial nucleosynthesis and microwave background data, it has
become clear that dark matter consists of nonbaryonic material.
There is a plethora of dark matter candidates.
Of the many candidates, the most popular are the weakly interacting massive particles
(WIMPS) and the axions, as these
particles have been proposed for other reasons in particle physics.
These are discussed further below.
Ordinary  neutrinos are too light to be cosmologically
significant, though sterile neutrinos remain a possibility.  Other
candidates include primordial black holes (for the latest bounds, see Ref.~\refcite{Carr:2016drx}),
self-interacting dark matter, light dark matter,
asymmetric dark matter, nonthermal WIMPzillas, Q-balls, and many others.

 \subsection{Axions}
\label{sec:axions}

The good news is that cosmologists don't need to ``invent'' new
particles.  Two candidates already exist in particle physics for other
reasons: axions and WIMPs.  Axions arise in the Peccei-Quinn solution to the strong-CP
problem in the theory of strong interactions, \cite{peccei} and are suitable dark matter candidates \cite{weinberg1978,wilczek1978} if the mass lies in the range $m_a \sim 10^{-(3-6)}\,$eV. An upper bound on the axion mass $m_a < 15\,$meV can be derived from astrophysical considerations,  \cite{raffelt1986, raffelt2008,viaux,miller_bertolami} while a lower bound comes from cosmology \cite{preskill1983,abbott1983,dine1983} and its value strongly depends on the thermal history of the universe and on the amount of topological defects. \cite{visinelli2010} An exclusion region $6 \times 10^{-13}{\rm\,eV} < m_a < 2 \times 10^{-11}{\rm \, eV}$ that is independent of the cosmological history and comes from black hole super-radiance has been obtained  \cite{arvanitaki} using aLIGO measurements.
Microwave cavity searches  \cite{sikivie1983} allow for a direct detection of axions. The Adark matterX cavity experiment  (ADMX)\cite{rosenberg} has already excluded a portion of the axion mass range and is currently searching for axions with a mass $\sim 10^{-5}\,$eV. A different technique consisting of searching for keV photons from axion-photon conversion in the Sun (through the Primakoff effect) has also been used in the KEK, CAST, and IAXO observatories. Such ``axion helioscopes'' are sensitive to the heavier end of the axion mass window.  In addition, new ideas for axion searches include the Cosmic Axion Spin Precession Experiment (CASPEr) \cite{Budker:2013hfa} and broadband and resonant approaches. \cite{Kahn:2016aff}
Axion searches continue to reach into the theoretically best motivated regions of mass and coupling.

\subsection{WIMPs}
\label{sec:WIMPs}

WIMPs are 
thought to be good dark matter candidates from particle physics for two reasons.
They are defined to be particles that participate in weak interactions (but not strong or electromagnetic)
and their masses are in the range GeV--10 TeV. 
These particles, if present in thermal abundance in the early
universe, annihilate with one another so that a predictable number of
them remain today.  The relic density of these particles comes out to
be the right value:
\begin{equation}
\Omega_\chi h^2 = (3 \times 10^{-27} {\rm cm}^3/{\rm sec})
/ \langle \sigma v \rangle_{ann}\,.
\end{equation}
Here $h$ is the Hubble constant in units of 100 km/s/Mpc, and
 the annihilation cross section $\langle \sigma v \rangle_{ann} $
of weak interaction strength automatically gives the correct abundance of these particles today.
This coincidence is known as ``the WIMP miracle" and is the first reason why
WIMPs are taken so seriously as dark matter candidates.

Secondly, WIMP candidates automatically exist in models that have been proposed to resolve problems
in theoretical particle physics. These models contain WIMPs as a byproduct of the theory. 
For example WIMP candidates exist in supersymmetric models (SUSY), including the lightest
neutralino in the minimal supersymmetric standard model.
Supersymmetry in particle theory is designed to keep particle masses
at the right value.  As a consequence, each particle we know has a
partner: the photino is the partner of the photon, the squark is the
quark's partner, and the selectron is the partner of the electron.
The lightest superysmmetric partner is a good dark matter candidate.
Another type of WIMP exists in models of universal extra dimensions.
In these theories all standard model fields propagate in a higher dimensional
bulk that is compactified on a space that is TeV$^{-1}$ in extent.  Higher
dimensional momentum conservation in the bulk translates in four dimensions
to Kaluza-Klein (KK) number (with boundary conditions to KK parity).  The lightest
KK particle, known as the LKP, does not decay and is a WIMP candidate. \cite{Servant:2002aq}
WIMP candidates are well-motivated from the point of view of particle physics and relic density;
the key issue now is whether or not nature agrees with our theoretical prejudice.
The experimental hunt for WIMPs is ongoing.

\section{Four Pronged Approach to WIMP Detection}

There are several ways to search for  WIMPs based on their interactions
with standard model particles: production at the Large Hadron Collider, scattering in underground direct detection experiments,
indirect detection of the products of annihilating dark matter, and discovery of dark stars. I will discuss each of these in turn.

\subsection{Production at the Large Hadron Collider at CERN}

At the Large Hadron Collider (LHC), protons are accelerated to 13 TeV.  Two beams travel in opposing directions
around a 27 kilometer long ring, and then collide in several detectors.  The two general purpose detectors ATLAS and CMS
were built with the goal of discovering the Higgs, discovering SUSY and dark matter, and discovering the unknown.
The first goal of finding the Higgs boson, the last missing piece of the standard model of particle physics,
 was successful as of July 2012 and immediately led to a Nobel Prize for
Higgs and Englert.  The other goals have as yet been elusive.

SUSY dark matter particles could manifest at the LHC in a variety of ways.
A possible signature would be missing transverse energy as the dark matter
particle leaves undetected, together with jets of particles created during the decay chain of SUSY particles emerging from the
collision.  Such a signature has not yet been seen, leading to ever higher bounds on SUSY particle masses.
The minimal supersymmetric standard model  (MSSM) has 105 free parameters.  If one makes some simplifying assumptions that
unify all fermion masses $m_{1/2}$ and all scalar masses $m_0$ at a high scale, then in the resulting constrained minimal
supersymmetric model (CMSSM, or MSUGRA), only five parameters remain.  The experimental results are often quoted in the context
of this CMSSM/MSUGRA.  For example, Fig.~6 illustrates the bounds from ATLAS on the supersymmetric parameter space.
The remaining parameter space is being pushed to above the TeV scale. However, it is important to note that
these bounds apply only to the MSUGRA/CMSSM.

\begin{figure}
\includegraphics[width=\textwidth]{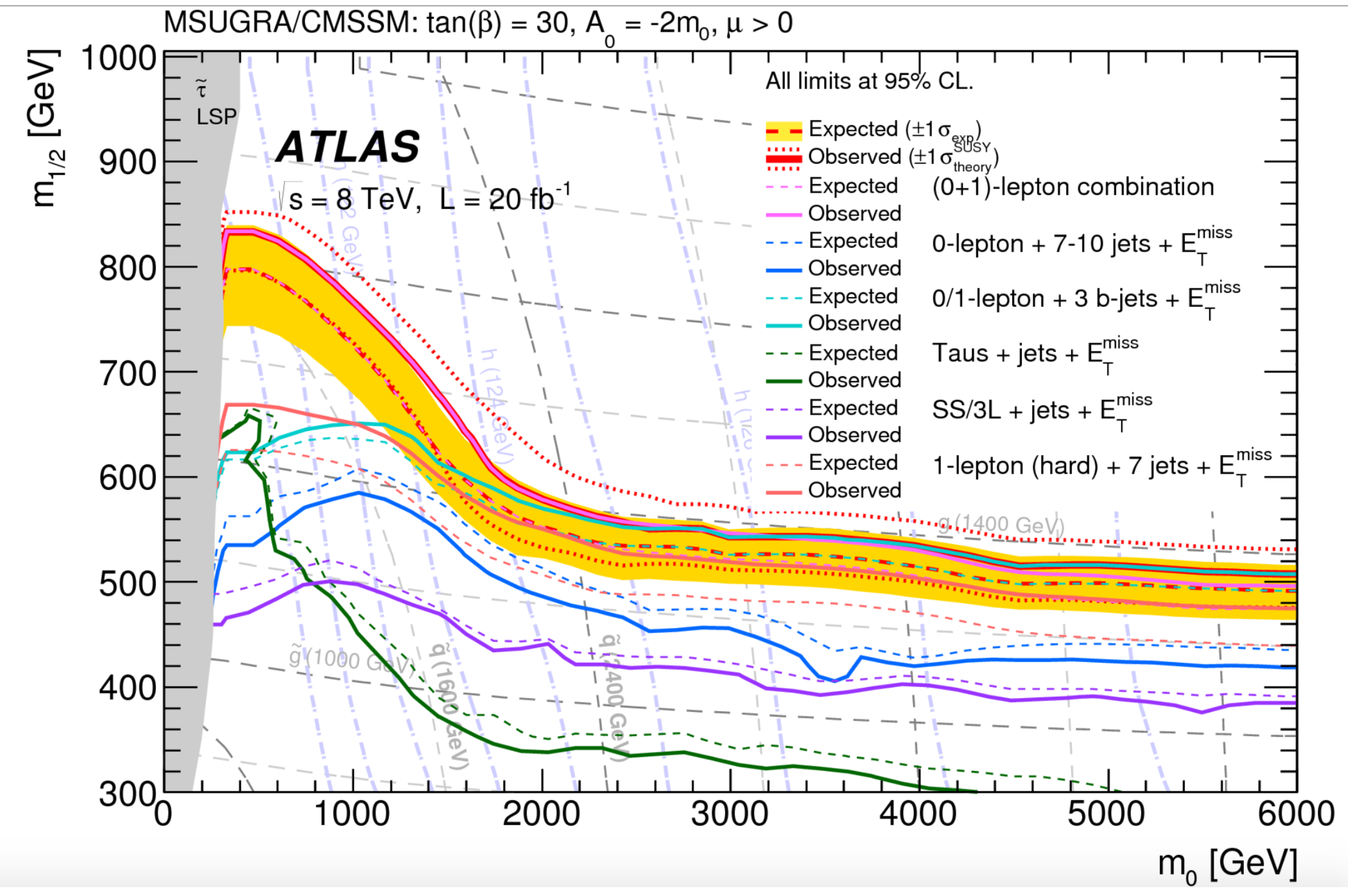}
\caption{Bounds on MSUGRA/CMSSM from 8 TeV  ATLAS data.  The remaining allowed parameter space is above the lines.}
\end{figure}

The LHC will never be able to kill even minimal supersymmetry. \cite{gates}  Even in the MSSM, a 25 GeV neutralino currently
survives as a possibility. \cite{Pierce:2013rda}  If the LHC sees nothing, SUSY can survive.  It may be at high scale.
Or, it may be less simple than the assumption that all scalars and all fermions unify at some high scale; e.g. the non-universal
Higgs model (NUHM) or the non-universal gaugino model (NUGM).

SUSY particles may be
discovered at the LHC as missing transverse energy plus jets in an event.  In that case one
knows that the particles live long enough to escape the detector, but
it will still be unclear whether they
are long-lived enough to be the dark matter.  Thus
complementary astrophysical experiments are needed.
Proof that the dark matter has been found requires astrophysical particles to be found,
via the other prongs of the dark matter search techniques.

\subsection{Direct Detection Experiments}

Direct detection experiments take advantage of the large number of WIMPs in the Galaxy.
A WIMP travels through the detector,
 scatters off of a nucleus, and deposits a small amount of energy that may be detected
 The experiments are extraordinarily difficult and the progress has been impressive:
the count rates are less than one count/kg/day and the energy deposited is O(keV).

The history of dark matter direct detection began with the ideas and theoretical calculations in the 1980s.
In 1984 Drukier and Stodolsky \cite{stodolsky} proposed neutrino detection via weak scattering off nuclei.
Then Goodman and Witten \cite{gw} turned the same approach to dark matter detection.
 Drukier, Freese, and Spergel \cite{dfs}
first  included a Maxwellian distribution of WIMPs in the Galaxy,
computed cross sections for a variety of candidates, and proposed the idea of annual modulation to identify
a WIMP signal.  In another paper we further studied the idea of using annual modulation, not only for background rejection but also to tease out a
signal even in the presence of overwhelming noise; \cite{wgould} this is the technique used by the DAMA experiment described below.
For  reviews, see Refs.~\refcite{Jungman_etal1996,jkgb,jkgc,Bertone_etal2004,lisanti}.

The text in the subsequent few paragraphs outlines dark matter direct detection and
 is taken from my review paper with Lisanti and Savage. \cite{lisanti}
When a WIMP strikes a nucleus, the nucleus recoils with energy $E$.
The differential recoil rate per unit detector mass is
\begin{equation}\label{eqn:dRdEnr}
 dR/dE
    = \frac{n_\chi}{M} \Big\langle v \frac{d\sigma}{dE} \; \Big\rangle
    = \frac{2\rho_\chi}{m_\chi}
      \int d^3v \, v f(v,t) \frac{d\sigma}{dq^2}(q^2,v) \, ,
\end{equation}
where $n_\chi = \rho_\chi/m_\chi$ is the number density of WIMPs, with
$\rho_\chi$ the local dark matter mass density; $f(v,t)$ is the
time-dependent WIMP velocity distribution; and
$\frac{d\sigma}{dq^2}(q^2,v)$ is the velocity-dependent differential
cross-section, with $q^2 = 2 M E$ the momentum exchange in the
scatter.  The differential rate is typically given in units of
cpd\,kg$^{-1}$\,keV$^{-1}$, where cpd is counts per day.  Using the
form of the differential cross-section for the most commonly assumed
couplings, to be discussed below,
\begin{equation}\label{eqn:dRdEnr2}
  dR/dE
    = \frac{1}{2 m_\chi \mu^2} \, \sigma(q) \, \rho_\chi \eta(v_{min}(E),t),
\end{equation}
where $\sigma(q)$ is an effective scattering cross-section and
\begin{equation} \label{eqn:eta}
  \eta(v_{min},t) = \int_{v > v_{min}} d^3v \, \frac{f(v,t)}{v}
\end{equation}
is the mean inverse speed, with
\begin{equation} \label{eqn:vmin}
  v_{min} =
      \sqrt{\frac{M E}{2\mu^2}}
       \end{equation}
The benefit of writing the recoil spectrum in the
form of Eqn.(\ref{eqn:dRdEnr2}) is that the particle physics and astrophysics
separate into two factors, $\sigma(q)$ and $\rho_\chi \eta(v_{min},t)$,
respectively.
It is traditional to define a form-factor
corrected cross-section
\begin{equation}\label{eqn:sigmaq}
  \sigma(q) \equiv \sigma_0 F^2(q) \, ,
\end{equation}
Here $\sigma_0$ is the scattering cross-section in the
zero-momentum-transfer limit and
$F^2(q)$ is a form factor to account for the finite size of the
nucleus.

Two types of interactions are most commonly studied.
In spin independent (SI) interactions,  the
scattering is coherent and scales as the atomic mass squared, $A^2$.
The SI
cross-section can be written as
\begin{equation} \label{eqn:sigmaSI2}
  \sigma_{SI} = \frac{\mu^2}{\mu_p^2} A^2 \, \sigma_{p,SI} \, ,
\end{equation}
where $\mu_p$ is the WIMP-proton reduced mass.
The SI cross-section grows rapidly with nuclear mass.  The explicit
$A^2$ factor arises from the fact that the
contributions to the total SI cross-section of a nucleus is a coherent
sum over the individual protons and neutrons within.

Spin dependent (SD) scattering is due to the interaction of a WIMP with the spin of the
nucleus and takes place only in those
detector isotopes with an unpaired proton and/or unpaired neutron.
The SD WIMP-nucleus cross-section is
\begin{equation} \label{eqn:sigmaSD}
  \sigma_{SD} = \frac{32 \mu^2}{\pi} G_{F}^{2} J(J+1) \Lambda^2 \, ,
\end{equation}
where $G_F$ is the Fermi constant, $J$ is the spin of the nucleus,
\begin{equation} \label{eqn:Lambda}
  \Lambda \equiv \frac{1}{J} \Big( a_p\langle S_p \rangle + a_n \langle S_n \rangle
 \Big) \, ,
\end{equation}
where $\langle S_p \rangle$ and $\langle S_n \rangle $
 are the average spin contributions from the
proton and neutron groups, respectively, and $a_p$ ($a_n$) are the
effective couplings to the proton (neutron) (these need not be the same).

The dark matter halo in the local neighborhood is most likely
dominated by a smooth and well-mixed (virialized) component with an
average density $\rho_\chi \approx 0.4$~GeV/cm$^3$.
The simplest model for this smooth component is often taken to be the
standard halo model (SHM) \cite{dfs,wgould} of an
isothermal sphere with an isotropic, Maxwellian velocity distribution
and rms velocity dispersion $\sigma_v$.  The SHM is written as
%
%
\begin{equation} \label{eqn:TruncMaxwellian}
  \widetilde{f}(v) =
      \frac{1}{N_{esc}} \left( \frac{3}{2 \pi \sigma_v^2} \right)^{3/2}
        \, e^{-3v^2\!/2\sigma_v^2} ,
         \textrm{for} \,\, |v| < v_{esc}  \\
         \end{equation}
and     $ \widetilde{f}(v) =  0 $ otherwise.  Here,
\begin{equation} \label{eqn:Nesc}
  N_{esc} = \textrm{erf}(z) - \frac{2}{\sqrt{\pi}} z e^{-z^2} \, ,
\end{equation}
with $z \equiv v_{esc}/v_0$, is a normalization factor and
\begin{equation} \label{eqn:vmp}
  v_0 = \sqrt{2/3} \, \sigma_v
\end{equation}
is the most probable speed, with an approximate value of 235~km/s
(see Refs.~\refcite{Kerr:1986hz,Reid:2009nj,McMillan:2009yr,Bovy:2009dr}).

Our early work \cite{dfs,wgould} used this Maxwellian dark matter distribution.  Although there has been concern
that the velocity distribution of the dark matter might deviate significantly from Maxwellian, Refs.~\refcite{withbaryons,bozorgnia,sloane2016} showed that results obtained for dark matter with a Maxwellian profile are consistent with those obtained when baryons are included in dark matter simulations, though there is as yet possible disagreement for the high velocity tail.  We concluded that
the Maxwellian approximation
 is a perfectly good approximation when comparing results of dark matter experiments to data.

 We also showed \cite{dfs} that the dark matter signal should experience an annual modulation
(for a review, see Ref.~\refcite{lisanti}.)
 As the Sun orbits around the Galactic Center, Earth-based detectors are effectively moving into a ``wind" of WIMPs.
 The WIMPs are moving in random directions in the Galaxy, and the Sun's motion creates (on the
 average) a relative velocity between us and the WIMPs.
  On top of that, because the Earth is moving around the Sun, the relative velocity of the Earth with the WIMP wind
 varies with the time of year. Thus the count rate should modulate sinusoidally with the time of year, peaking in June
 and with a minimum in December.
   We predicted that the annually modulating
recoil rate can be approximated by
\begin{equation} \label{eqn:dRdES}
  dR/dE(E,t) \approx S_0(E) + S_m(E) \cos{\omega(t-t_0)} ,
\end{equation}
with $|S_m| \ll S_0$, where $S_0$ is the time-averaged rate, $S_m$ is
referred to as the modulation amplitude, $\omega = 2\pi$/year and $t_0$ is the phase of the
modulation.
Since typical backgrounds do not experience the same annual modulation, this effect
can be used to tease the signal out of the background. \cite{wgould}

 These first papers convinced experimentalists that they would be able to build detectors sensitive
 enough to search for WIMPs. The detectors must be placed deep underground in order to filter out cosmic rays,
 in underground mines or underneath mountains.  The first experimental effort to search for and bound WIMP dark
 matter was Ref.~\refcite{ahlenavignone}.
 Now, 30 years later, direct detection searches are
 ongoing worldwide, in US, Canada, Europe, Asia, and the South Pole, see Fig.~7.

\begin{figure}
\includegraphics[width=\textwidth]{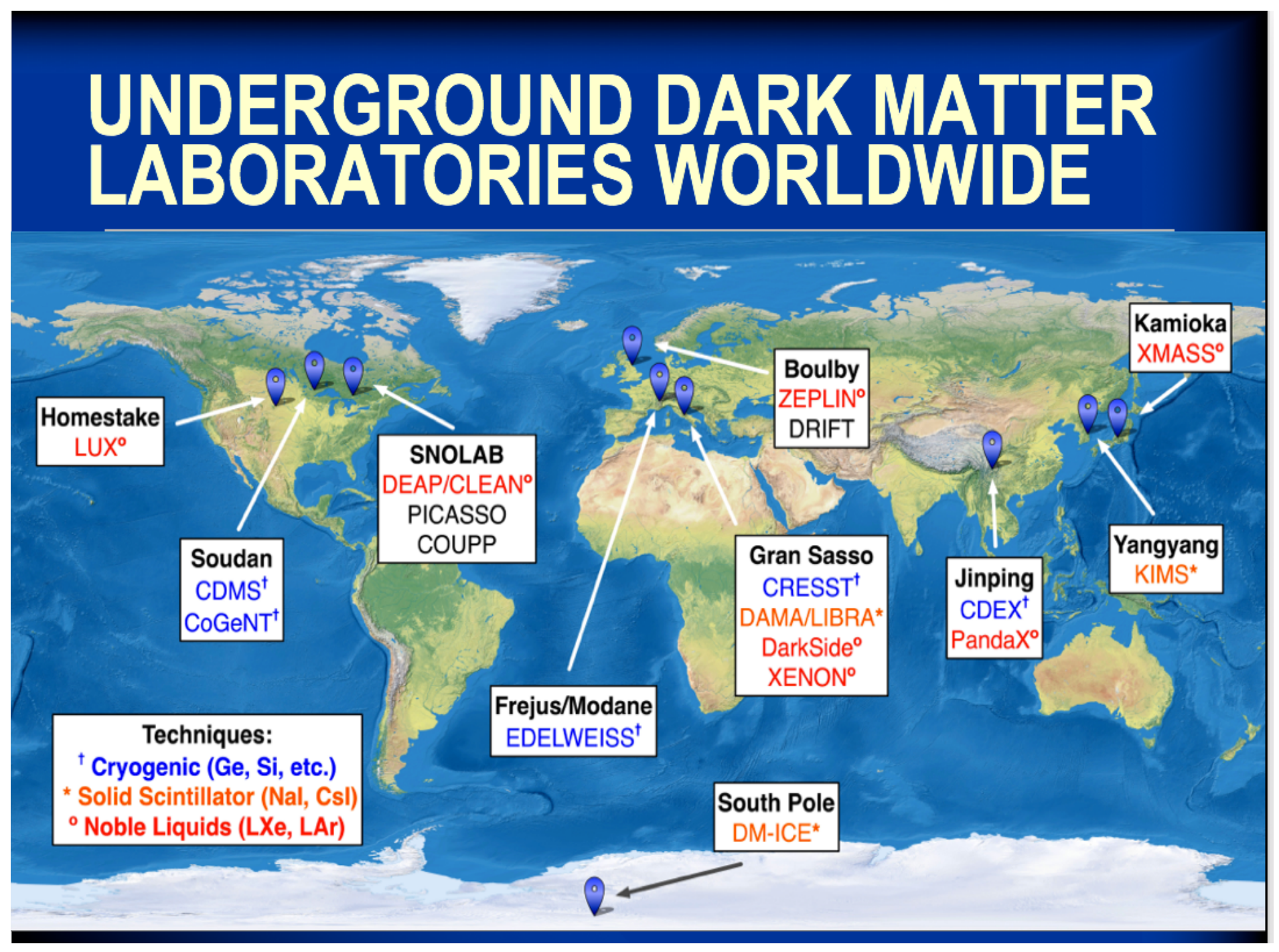}
\caption{Underground dark matter laboratories worldwide (courtesy of M. Tripathi and M. Woods). The CanFranc underground
laboratory in Spain is missing from the figure.}
\end{figure}

 Of all of these experiments, only one, the Italian DAMA experiment, \cite{Bernabei:2014tea} has positive signal.
 They use NaI crystals in the Gran Sasso tunnel under the Apennine Mountains near Rome.
 The signal they have is the annual modulation we predicted for the WIMP signal. \cite{dfs,wgould}
 DAMA has observed exactly this annual modulation with the correct phase, see Fig.~8. 
 Indeed DAMA has 10 years of cycles corresponding to a 9 $\sigma$ detection of modulation.

 Now the question is, have they detected dark matter? Unfortunately they have not released the data for others to study.
 In addition, no experiment other than DAMA has found any signal at all.  Indeed the null results from other experiments
 place strong bounds on the WIMP elastic scattering cross section.  Naively it might seem that the other experiments
 rule out the DAMA results as being due to WIMPs. Yet, this may not be true, because all the detectors are made of
 different materials.  DAMA is the only experiment to date that uses NaI crystals.
  For example, LUX \cite{Akerib:2016vxi}
and XENON \cite{Aprile:2016swn}
are made of xenon while CDMS (and SuperCDMS) \cite{cdms} is made of germanium, which are far heavier nuclei
 than the components of DAMA's NaI crystals.  To compare the different experiments, theoretical input is required.
 For example, if one assumes the scattering is SI so that the cross section scales as $A^2$, one can then plot
 the different experiments as in Fig.~9 in the cross section/ WIMP mass plane.  DAMA signals
could  be due to roughly 10 GeV WIMPs if the scattering is with Na atoms, while the signal would be due to 80 GeV WIMPs
if the scattering were off of iodine atoms.  The higher mass region is in severe conflict with bounds from other
experiments, while the lower mass region also appears to be ruled out.  However, if one abandons the $A^2$ assumption
then this comparison plot is no longer valid.  For all known theoretical assumptions it is hard to reconcile the positive results
of DAMA with the negative results of other experiments.  Perhaps uncertain nuclear physics may be responsible.
\cite{Anand:2014kea,Dent:2015zpa}
 Many alternate explanations to the discovery of DM
have been proposed (e.g. radon contamination, muons, etc.) but all have been shown to be wrong.
The reason DAMA remains so interesting is that there is no other known explanation of the
annual modulation they are seeing. 

 \begin{figure}
\includegraphics[width=\textwidth]{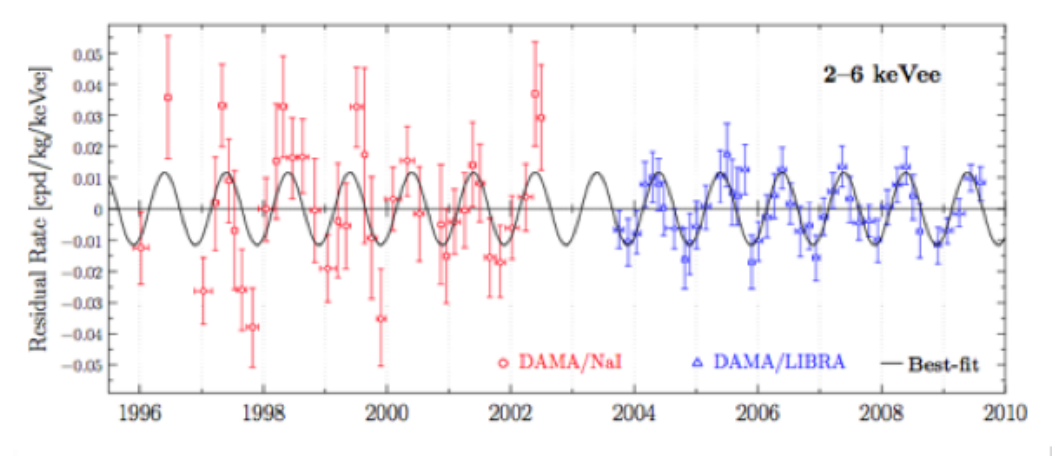}
\caption{DAMA data (including DAMA/LIBRA) has a 9 $\sigma$ detection of annual modulation consistent
with WIMPs. \cite{Bernabei:2014tea}}
\end{figure}

What is needed are further experimental tests using the same detector material as DAMA (NaI crystals) but in a different
location.  These experiments are now taking place:  SABRE, \cite{Froborg:2016ova}
COSINE-100 \cite{COSINE100} (KIMS has joined with dark matter-ICE \cite{deSouza:2016fxg}), and ANAIS. \cite{Amare:2015rpa}
 Thus in the next five years there should be either
confirmation of DAMA or it will be ruled out.

 \begin{figure}
\includegraphics[width=\textwidth]{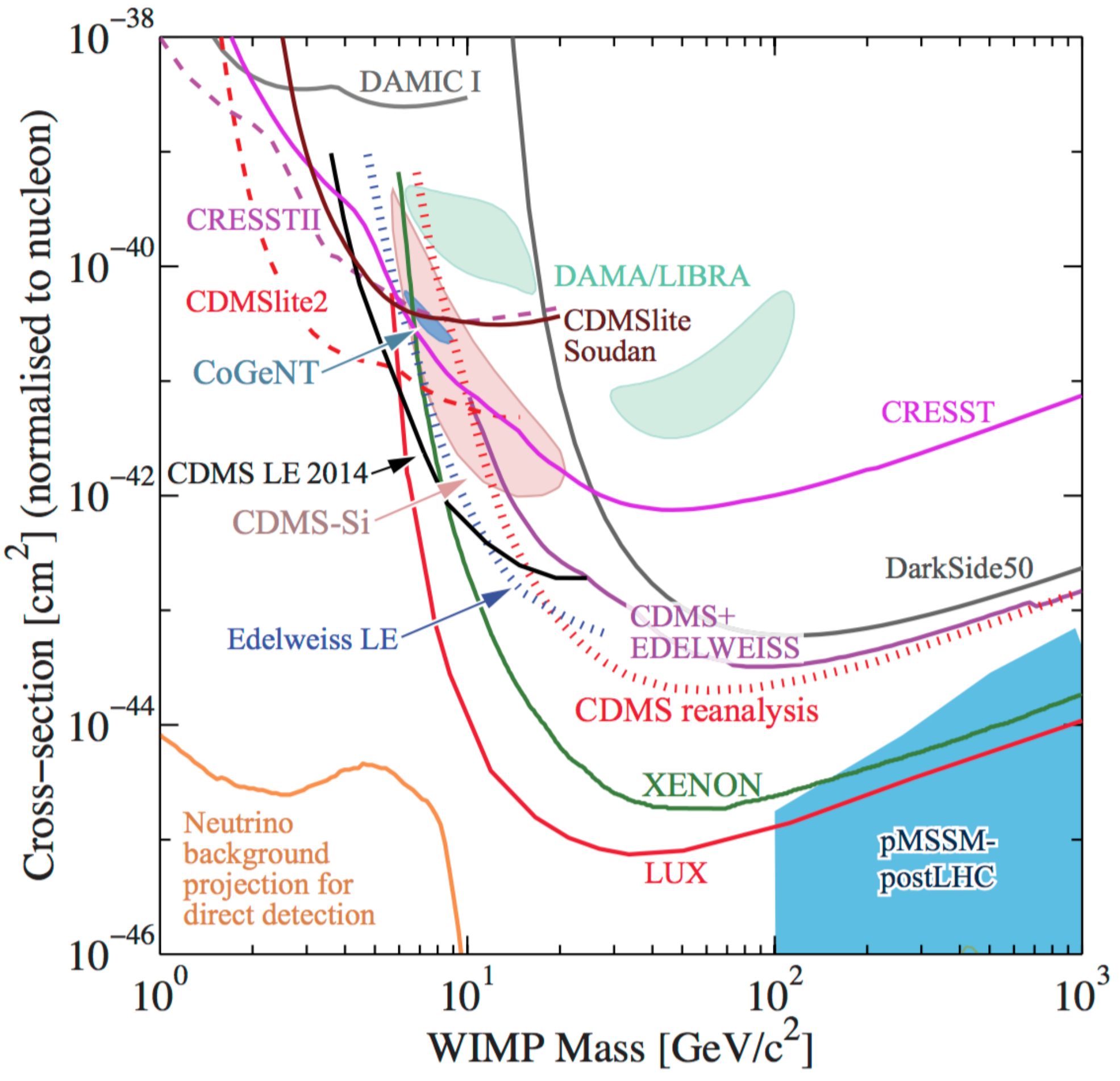}
\caption{Spin independent scattering bounds from direct detection experiments as shown, as well as regions compatible with DAMA data,
in the SI elastic scattering cross section vs. WIMP mass plane. Plot taken from Particle Data Book 2015 (PANDA-X and LUX bounds need to be updated).}
\end{figure}

I also wanted to mention a new idea we have for dark matter direct detection using DNA (see Fig.~10). We proposed \cite{DNA} to
use nanometer thin sheets of gold (or other material) with $\sim10^{60}$ strands of DNA attached.  When a WIMP hits the
gold sheet, it knocks a gold atom forward into the DNA.  The gold atom then severs whatever DNA strands it hits.
The broken strand of DNA then falls down and is collected.
The DNA has been carefully constructed to have a well-known sequence of bases (A,G,C...).
Using well known biological techniques (PCR and sequencing), the location of the break can be identified.
Thus the track of the recoiling gold nucleus can be reconstructed.  Since the distance between the bases in the DNA
strand is nanometer in size, this technique provides a nanometer tracker.
Once the track of the gold nucleus is known, since the WIMP traveled in roughly the same direction, the direction that
the WIMP came in from is also known.  This idea thus provides a directional dark matter detector.   The importance of this
is as follows.  It allows clear proof of dark matter discovery.  We expect ten times as many counts when the detector
is pointing into the direction of the WIMP wind than when it is pointing in the opposite direction.  This head/tail asymmetry
would be hard to explain with any background.  Additionally, both the annual and daily (due to Earth's rotation)
 variation of the signal would
be detectable and would give superb background rejection.  In the long run, a directional detector would allow
the discovery of where the WIMPs are in the Galaxy and how they are moving.

A second radically new idea we have proposed for dark matter detection is ``nanobooms". \cite{nanobooms}
The WIMP sets off a very small explosion when
it deposits heat in the detector.  For example, the detector might consist of thermites. Then the WIMP's energy deposit would
cause the exothermic reaction between a metal and a metal oxide to take place, i.e., there is a small explosion, which can then
be detected acoustically, optically, or more likely via gas expansion.

 \begin{figure}
\includegraphics[width=\textwidth]{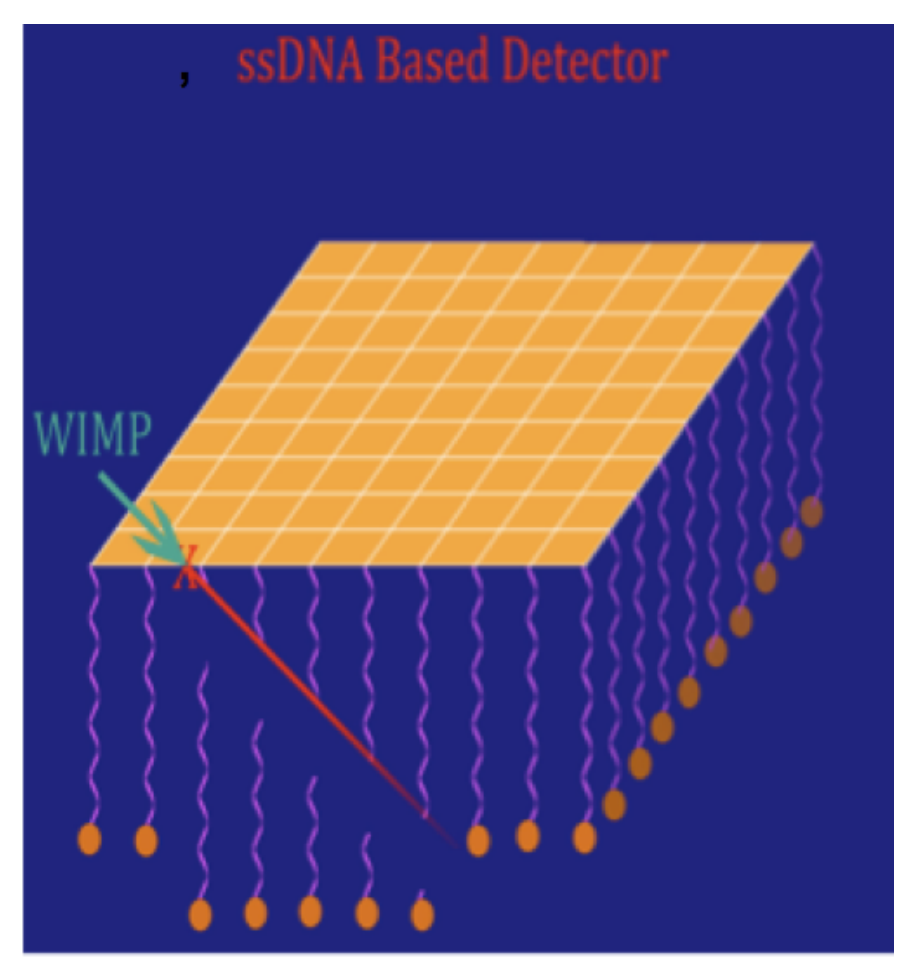}
\caption{DNA based Dark Matter Directional Detector.
A WIMP hits the nanometer-thin gold plate, knocks a gold atom into the hanging
strands of DNA. Whenever the gold atom strikes a DNA strand, the strand breaks and is collected.  Since the base sequence
of the strands is controlled, sequencing the broken strand allow the location of the break to be identified.
Hence the DNA serves as a tracker
with nanometer accuracy.  Since the WIMP travels in roughly the same direction as the gold atom, the detector discovers the
direction the WIMP came from.}
\end{figure}

The next five years stand to lead to tests of the DAMA annual modulation signal and a confirmation or refutation of WIMP
discovery as well as progress in directional sensitivity.

 \subsection{Indirect Detection}
WIMP annihilation in today's universe takes place wherever there is an overdensity of WIMPs.
The final products of WIMP annihilation are neutrinos, e$^+$/e$^-$ pairs, and photons.
All three of these are being looked for in detectors.
Promising places to look are the Galactic Center, dwarf galaxies, clusters of galaxies, \cite{Adams:2016alz}
and in the case of neutrinos, the Earth and the Sun. The first papers suggesting
the latter neutrino searches were by Silk et al.\, \cite{SOS} in the Sun; and
by Freese \cite{Freese1986} as well as Krauss, Srednicki and
Wilczek \cite{Krauss_etal1986} in the Earth.  As yet no signal of neutrinos due to WIMP annihilation
in the Sun or Earth \cite{Aartsen:2016fep}
has been found
in the IceCube/DeepCore detectors at the South Pole.

The AMS experiment on board the International Space Station has found an excess of positrons. \cite{AMS}
 However, this excess is not likely to be due to WIMP annihilation.
A combination of two papers has shown that such an explanation is extremely unlikely.
First, the work of Lopez, Savage, Spolyar, and Adams \cite{Lopez:2015uma}
pointed out that
such a positron excess would predict also gamma-rays from dwarf galaxies, which are not seen in the Fermi Gamma Ray
Space Telescope (Fermi-LAT) data.
They used the bounds on gamma-rays from dwarfs in Fermi-LAT data 
to show that all WIMP annihilation channels are excluded as explanations of AMS data except one (via  a mediator to four muons).
This latter channel was further examined by Scaffidi et al.\ \cite{Scaffidi:2016ind}
Second, the Planck satellite
examined the effects such an excess would imply for the CMB and ruled out a large swath of parameter space. \cite{Ade:2015xua}
The work of  Ref.~\refcite{Lopez:2015uma} using Fermi-LAT data to rule out a DM explanation of the AMS positron excess
was placed on the arXiv a month prior to the Planck bounds.
It is far more likely that the AMS positron excess is due to pulsars or other point sources than due to WIMP
annihilation.

Of great interest over the last few years has been Fermi-LAT's discovery of a gamma-ray excess towards the Galactic Center.
Hooper and Goodenough \cite{GCexcess}  pointed out that it could be from the annihilation of a 40 GeV WIMP.
More recent studies of cosmic ray backgrounds have widened the possible range of masses \cite{agrawal}
and therefore SUSY
explanations of this excess.  \cite{Freese:2015ysa}
However, studies \cite{pointsources} have shown that  a point source explanation (e.g., pulsars) is at least as likely as
a dark matter explanation.  Though tantalizing, a dark matter explanation of this gamma ray excess will be hard to prove as there
is much astrophysical competition at the Galactic Center.

\subsection{Summary of WIMP Searches}

To summarize the current status of WIMP searches, there is possible evidence for WIMP detection already  now, but
none of it is certain.
The direct detection experiment DAMA has found annual modulation of its signal that would be compatible
with a WIMP origin. However, other experiments have null results in conflict with DAMA's result.  Since the
experiments are made of different detector materials, further tests of the same material as DAMA are now
taking place around the world and will result in confirmation or refutation in the next five years.

As far as indirect detection of WIMP annihilation products, the positron excess seen by AMS likely has a different
origin than WIMPs.  The gamma-ray excess seen from the direction of the Galactic Center by the Fermi Gamma Ray Space Telescope
is compatible with a WIMP origin but other astrophysical explanations are at least as likely.

Theorists are looking for models in which some of these results are consistent with one another, given a WIMP interpretation.
What will it take for us to believe dark matter has been found? We need a compatible signal in a variety of experiments made
of different detector materials and all the parties agree.

\section{Dark Stars}
A fourth prong of the hunt for dark matter is the search to discover dark stars.
The first stars to form in the universe, at redshifts $z \sim 10-50$,
may be very unusual; these dark stars are made almost entirely of atomic matter (hydrogen and helium, with only $10^{-3}$ of the mass made of dark matter)
and yet are powered by dark matter heating rather than by fusion.   Dark stars were first proposed by Spolyar, Freese, and Gondolo \cite{SpolyarFreeseGondolo08} and are reviewed in Ref.~\refcite{Freese:2015mta}.

 As discussed in the last section, WIMP dark matter annihilation in the
early universe provides the right abundance today to explain the dark
matter content of our universe. This same annihilation process will
take place at later epochs in the universe wherever the dark matter
density is sufficiently high to provide rapid annihilation.  The first
stars to form in the universe are a natural place to look for
significant amounts of dark matter annihilation, because they form at
the right place and the right time. They form at high redshifts, when
the universe was still substantially denser than it is today, and at
the high density centers of dark matter haloes.

The first stars form inside dark matter haloes of $\sim10^6 M_\odot$
(for reviews see e.g.,  Ripamonti \& Abel, \cite{RipamontiAbel05}
Barkana \& Loeb, \cite{BarkanaLoeb01} and Bromm \& Larson;
\cite{BrommLarson03} see also Yoshida et al.\ \cite{Yoshida_etal06}).
One star is thought to form inside one such dark matter halo.   It was our idea to ask, what is the
effect of the dark matter on these first stars?  We studied the behavior of
WIMPs in the first stars.  As our canonical values, we take $m_\chi =
100$GeV for the WIMP mass and $\langle \sigma v \rangle_{ann} = 3
\times 10^{-26} {\rm cm^3/sec}$ for the annihilation cross section
(motivated above).  However, the same behavior results for a wide variety
of WIMP masses and cross sections over many orders of magnitude.
We find that the annihilation products of the
dark matter inside the star can be trapped and
deposit enough energy to heat the star and prevent it from further
collapse.  A new stellar phase results, a dark star, powered
by dark matter annihilation as long as there is dark matter fuel.

\subsection{Three Criteria for Dark Matter Heating}

 WIMP annihilation produces energy at a rate per
unit volume
\begin{equation}
  Q_{\rm ann} = \langle \sigma v \rangle_{ann} \rho_\chi^2/m_\chi
  \linebreak \simeq  10^{-29} {{\rm erg} \over {\rm cm^3/s}} \,\,\, {\langle
    \sigma v \rangle \over (3 \times 10^{-26} {\rm cm^3/s})} \left({n \over {\rm
        cm^{-3}}}\right)^{1.6} \left({100 {\rm GeV}\over m_\chi}\right)
\end{equation}
where $\rho_\chi$ is the dark matter energy density inside the star and $n$ is
the stellar hydrogen density.  Spolyar, Freese and Gondolo
\cite{SpolyarFreeseGondolo08} outlined the three key ingredients
for dark stars: 1) high dark matter densities, 2) the annihilation
products get stuck inside the star, and 3) dark matter heating wins over other
cooling or heating mechanisms.  These  ingredients are required
throughout the evolution of the dark stars.

{\bf First criterion: high dark matter density inside the star.}  Dark
matter annihilation is a powerful energy source in these first stars
because the dark matter density is high. To find the dark matter density
profile, we started with an NFW (Navarro, Frenk \& White
\cite{NavarroFrenkWhite96}) profile for both dark matter and gas in the $10^6
M_\odot$ halo.  However, we find the same behavior results for even
a completely flat profile; the dark star is born regardless.
Originally we used adiabatic contraction ($M(r)r$ =
constant) (Blumenthal et al.\ \cite{Blumenthal_etal85}) and
matched onto the baryon density profiles given by Abel, Bryan \&
Norman \cite{AbelBryanNorman02} and Gao et
al.\ \cite{Gao_etal07} to obtain dark matter profiles; see also Natarajan,
Tan \& O'Shea \cite{NatarajanTanO'Shea08} for a recent
discussion.  Subsequent to our original work, we have done an exact
calculation (which includes radial orbits) (Freese, Gondolo, Sellwood
\& Spolyar \cite{FreeseGondoloSellwoodSpolyar08}) and found that
our original results were remarkably accurate, to within a factor of
two.  At later stages, we also consider possible further enhancements
due to capture of dark matter into the star (discussed below).

{\bf Second Criterion: dark matter annihilation products get stuck
  inside the star}.  In the early stages of Population III star formation,
when the gas density is low, most of the annihilation energy is
radiated away (Ripamonti, Mapelli \& Ferrara
\cite{RipamontiMapelliFerrara06}). However, as the gas collapses
and its density increases, a substantial fraction $f_Q$ of the
annihilation energy is deposited into the gas, heating it up at a rate
$f_Q Q_{\rm ann}$ per unit volume.  While neutrinos escape from the
cloud without depositing an appreciable amount of energy, electrons
and photons can transmit energy to the core.  We have computed
estimates of this fraction $f_Q$ as the core becomes more dense. Once
$n\sim 10^{11} {\rm cm}^{-3}$ (for 100 GeV WIMPs), e$^-$ and photons
are trapped and we can take $f_Q \sim 2/3$.

{\bf Third Criterion: dark matter heating is the dominant heating/cooling
  mechanism in the star}.  We find that, for WIMP mass $m_\chi =
100$GeV (1 GeV), a crucial transition takes place when the gas density
reaches $n> 10^{13} {\rm cm}^{-3}$ ($n>10^9 {\rm cm}^{-3}$).  Above
this density, dark matter heating dominates over all relevant cooling
mechanisms, the most important being H$_2$ cooling (Hollenbach
\& McKee \cite{HollenbachMcKee79}).

Fig.~11 shows evolutionary tracks of the protostar in the
temperature-density phase plane with dark matter heating included
(Yoshida et al.\ \cite{Yoshida_etal08}), for two dark matter particle
masses (10 GeV and 100 GeV).  Moving to the right on this plot is
equivalent to moving forward in time.  Once the black dots are
reached, dark matter heating dominates over cooling inside the star.
The protostar collapses somewhat further until it reaches equilibrium, at which point the
dark star phase begins.  The protostellar core is prevented from
cooling and collapsing further.  The size of the core at this point is
$\sim 17$ A.U. and its mass is $\sim 1 M_\odot$ for 100 GeV mass
WIMPs.  A new type of object is created, a dark star supported by dark matter
annihilation rather than fusion.

\begin{figure}
\includegraphics[width=\textwidth]{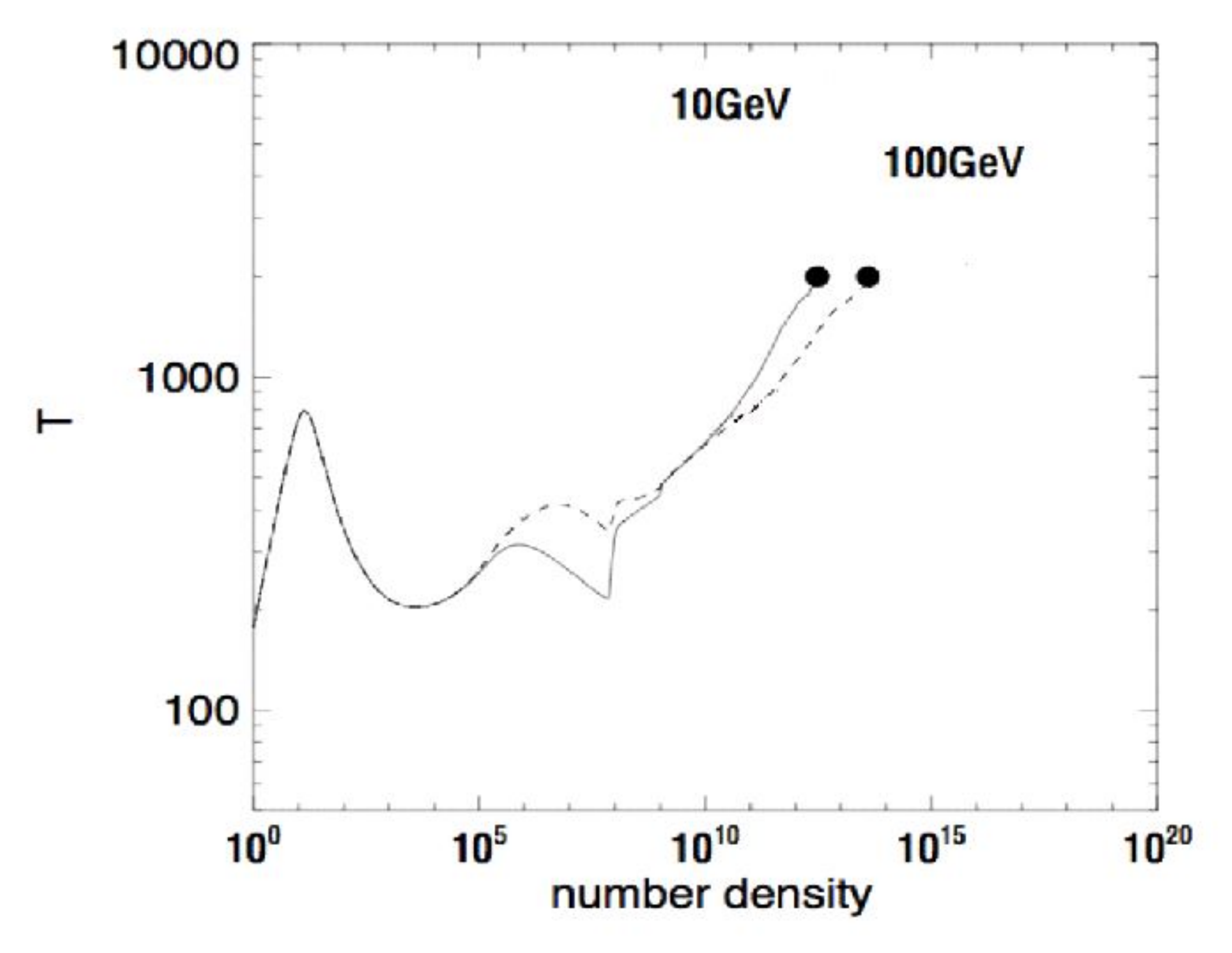}
\caption{ Temperature (in degrees K) as a function of hydrogen density
  (in cm$^{-3}$) for the first protostars, with dark matter annihilation
  included, for two different dark matter particle masses (10 GeV and 100 GeV).
  Moving to the right in the figure corresponds to moving forward in
  time.  When the ``dots'' are reached, dark matter annihilation wins over H2
  cooling.  After that the protostar collapses somewhat further until
  it reaches equilibrium.  At that point a dark star is created.}
\end{figure}

\subsection{Building up the Mass}

We have found the stellar structure of the dark stars
(hereafter DS) (Freese, Bodenheimer, Spolyar \& Gondolo
\cite{FreeseBodenheimerSpolyarGondolo08}).  They accrete mass from the
surrounding medium.  We built up the DS mass as it grows
from $\sim 1 M_\odot$ to possibly become supermassive.
The studies were done in two different ways, first assuming polytropic interiors and
more recently using the MESA stellar evolution code; the basic results are the same.
\cite{Rindler-Daller:2014uja}
As the mass increases, the DS radius adjusts
until the dark matter heating matches its radiated luminosity.  We find
solutions for dark stars in hydrostatic and thermal
equilibrium. We build up the DS by accreting $1 M_\odot$ at a time
with a variety of possible accretion rates, always
finding equilibrium solutions.  We find that initially the DS are in
convective equilibrium; from $(100-400) M_\odot$ there is a transition
to radiative; and heavier DS are radiative.  As the DS grows, it pulls
in more dark matter, which then annihilates.  Fig.~12 shows the hydrogen and dark matter density profiles. One can see ``the power of
darkness'': although the dark matter constitutes a tiny fraction ($<10^{-3}$)
of the mass of the DS, it can power the star. The reason is that WIMP
annihilation is a very efficient power source: 2/3 of the initial
energy of the WIMPs is converted into useful energy for the star,
whereas only 1\% of baryonic rest mass energy is useful to a star via
fusion.

\begin{figure}
\centering
\includegraphics[width=0.5\textwidth]{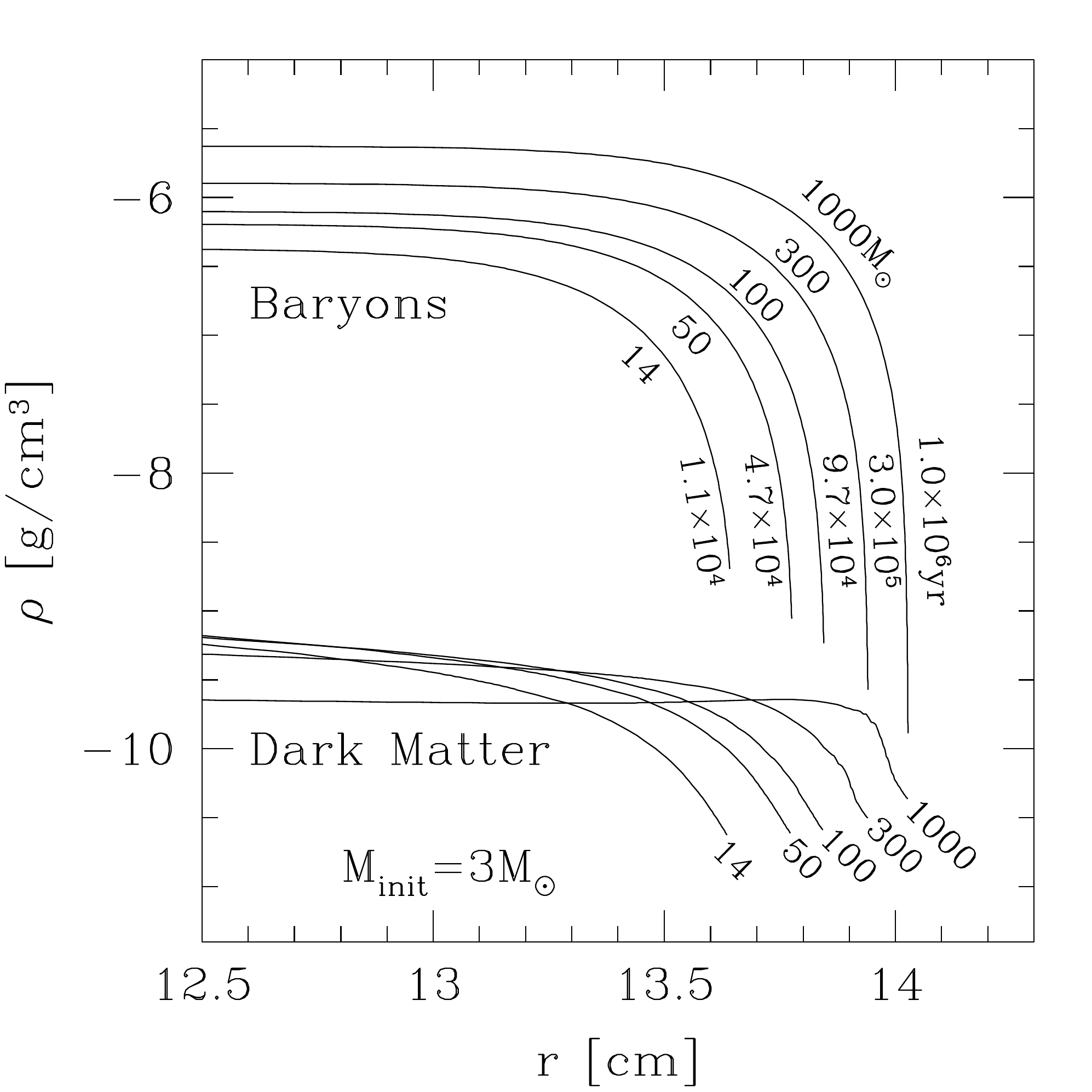}
\caption{Evolution of a dark star as mass is accreted onto the
  initial protostellar core of 3 M$_\odot$.  The set of upper (lower)
  curves correspond to the baryonic (dark matter) density profile at different
  masses and times. Note that dark matter constitutes $<10^{-3}$ of the mass of
  the DS.}
\end{figure}

\subsection{Later stages: Capture}

The dark stars will last as long as the dark matter fuel inside them persists.
Once the gravitationally attracted dark matter runs out, the star collapses somewhat,
at which point the star is dense enough to capture more dark matter.

The new source of dark matter in the first stars is capture of dark matter particles
from the ambient medium.  Any dark matter particle that passes through the
DS has some probability of interacting with a nucleus in the star
and being captured. The new particle physics ingredient required
here is a significant scattering cross section between the WIMPs
and nuclei. Whereas the annihilation cross section is
fixed by the relic density, the scattering cross section is a
somewhat free parameter, set only by bounds from direct detection
experiments.
Two simultaneous papers (Freese, Spolyar \& Aguirre,
\cite{FreeseSpolyarAguirre08} Iocco \cite{Iocco08}) found the
same basic idea: the dark matter luminosity from captured WIMPs can be larger
than fusion for the DS. Two uncertainties exist here: the scattering
cross section, and the amount of dark matter in the ambient medium to capture
from.  DS studies following the original papers that include
capture have assumed (i)  the maximal scattering cross sections allowed by
experimental bounds and (ii) ambient dark matter densities that are never depleted.
With these assumptions, DS evolution models with dark matter heating after the
onset of fusion were studied in several papers.  \cite{Taoso_etal08,Yoon_etal08}

\subsection{Supermassive Dark Stars}

Dark stars are very unusual stars --- they are made of atomic matter (hydrogen and helium)
but they are powered by dark matter heating (Freese, Bodenheimer, Spolyar \& Gondolo
\cite{FreeseBodenheimerSpolyarGondolo08}). They are very puffy (10 A.U. in size) and cool
(surface temperatures $\sim$ 10,000 K.  Reionization
during this period is likely to be slowed down, as these stars can
heat the surroundings but not ionize them.
Because they are so cool, they can keep accreting matter and growing
as long as there is dark matter fuel.
Standard Population III stars are hot, give off ionizing photons, and prevent further accretion
above $\sim 140 M_\odot$.  Dark stars, on the other hand, can keep growing to become supermassive,
even as massive as $10^7 M_\odot$ and as bright as $10^{10} L_\odot$.
There should be a variety of dark star masses ranging from a few solar masses all the way up to these
very large masses.

Fig.~13 shows the Hertzsprung-Russell  diagram for dark stars as they grow from $\sim 1 M_\odot$ to become supermassive.
The two cases of matter being accreted gravitationally and via capture are shown separately. 

\begin{figure}
\centering
\includegraphics[width=0.5\textwidth]{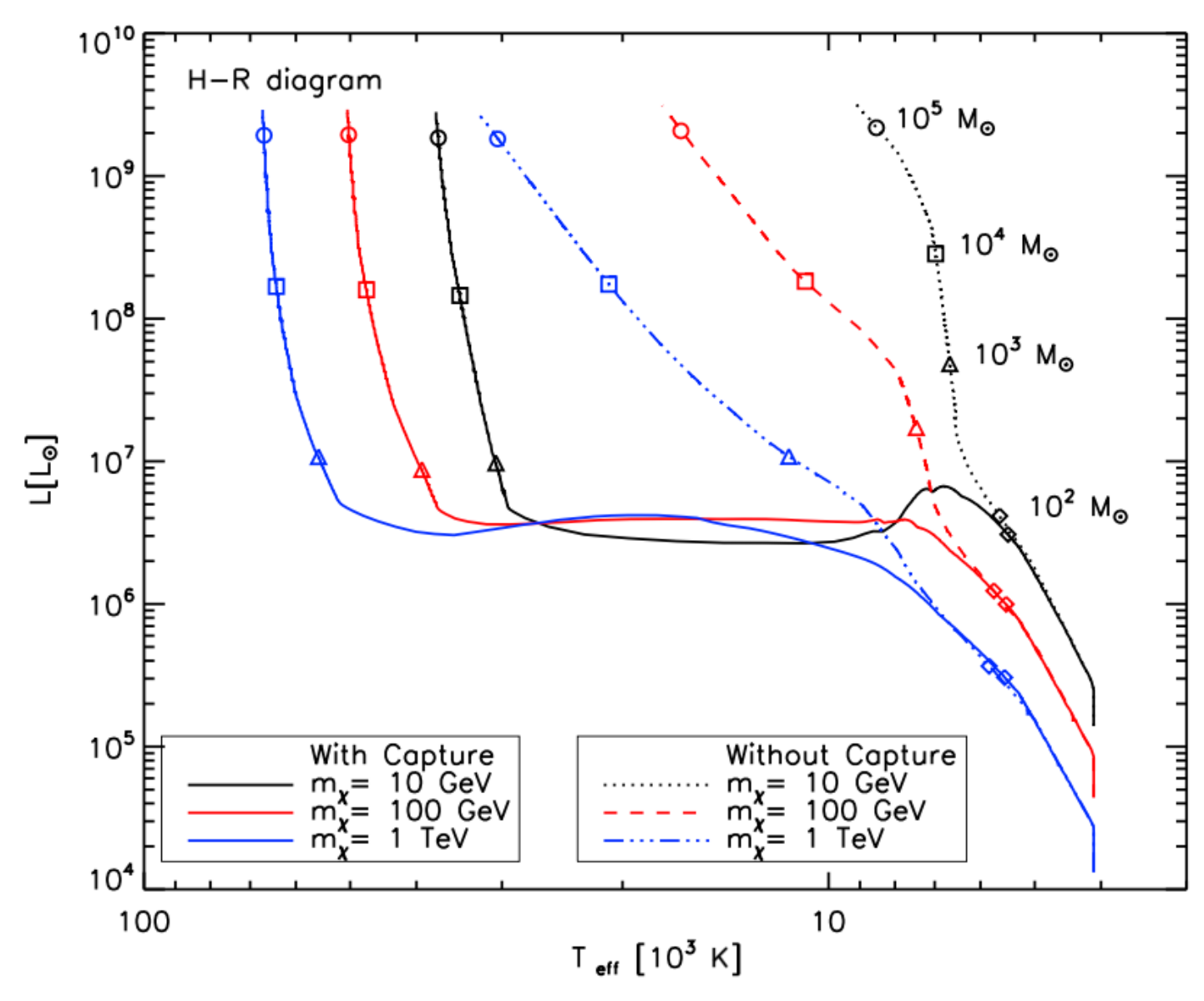}
\caption{Hertzsprung-Russell diagram for DSs for a variety of WIMP masses as labeled for the two cases: (i) with gravitationally
attracted dark matter only (dotted lines), assuming no significant depletion of dark matter due to annihilation, which is equivalent to assuming a replenishment of dark matter due to centrophilic orbits; (ii) with capture (solid lines). Results were obtained assuming polytropic interiors for the DS. The case with capture is for product of scattering cross section times ambient WIMP density
$\sigma_c \rho_\chi = 10^{-39}$ GeV/cm$^3$ (the maximum allowed cross section for all WIMP masses and the maximum reasonable ambient density for 100 GeV WIMPs).  Once the gravitational dark matter runs out,  DSs must first become dense enough in order
for dark matter capture to happen. This explains the horizontal lines in the evolution of the case with capture. Labeled are also stellar masses reached by the DS on its way to becoming supermassive. The final DS mass was taken to be $10^5 M_\odot$ (the baryonic mass inside the initial halo), but could vary from halo to halo, depending on the specifics of the halo mergers (figure taken from Ref.~\protect\refcite{HR}).
}
\end{figure}

\subsection{Dark Stars are Detectable in James Webb Space Telescope}
Supermassive dark stars may be detectable in the JWST as J, H, or K-band dropouts.  Detailed discussion may be found
in Refs.~\refcite{HR,ruiz,cosmin}.  Comparison of light output with sensitivity of JWST filters is shown in Fig.~14
for a $10^6 M_\odot$ DS.  Predictions for numbers of these objects, based on cosmological simulations, is also found in Ref.~\refcite{HR}.

\begin{figure}
\centering
\includegraphics[width=0.5\textwidth]{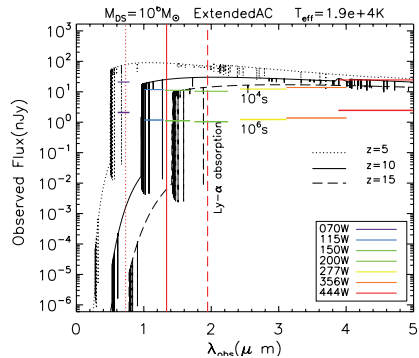}
\caption{Supermassive Dark Stars in JWST.  Spectra for $10^6 M_\odot$ supermassive DSs formed at  redshift $z_{form}=15$ compared with sensitivity of JWST filters. The formation mechanism in this figure is gravitational attraction of dark matter only. The surface temperature $T_{\rm eff} = 1.9 \times 10^4$K. The fluxes are shown at z = 15 (dashed line), 10 (solid line) and 5 (dotted line) and compared to the detection limits of NirCam wide passband filters. The colored horizontal lines represent the sensitivity limits for the filters as labeled in the legend for exposure times $10^4$ sec (upper lines) and $10^6$ sec (lower lines). IGM absorption will decrease the observed fluxes for wavelengths shortward of the vertical red lines, which indicate the Lyman-$\alpha$ line (1216 Angstroms) redshifted from the rest-frame of the star (figure taken from Ref.~\protect\refcite{cosmin}).}
\end{figure}

\subsection{Supermassive Black Holes}
Once these supermassive dark stars (SMDS) run out of dark matter fuel, they collapse to black holes.
They may provide large seeds for the supermassive black holes that have been found
at high redshift ($10^9-10^{10} M_\odot$ BH at $z=6$) and are, as yet,
unexplained (Li et al., \cite{Li_etal07} Pelupessy et
al., \cite{Pelupessy_etal07} Wu et al.\ \cite{WuNature}).

\subsection{Pulsations} 

An interesting new research direction is the fact that DS pulsate, like all stars.  As a first step, we used the MESA stellar
evolution code to calculate  the adiabatic pulsation periods of radial $p$-modes (where the
restoring force is pressure and those for which there is no angular dependence, so $l = 0$).
We found that our DS models pulsate on timescales which range from less than a day to more than two years in their restframes at about $z = 15$, depending on the WIMP mass and overtone number. The pulsation periods are significantly shorter for higher WIMP mass. Converting to the observer frame, the shortest periods we found are less than about 50 days for modes with overtone number
$n > 6$ and a WIMP mass of 1 TeV (Ref.~\refcite{Rindler-Daller:2014uja}).
We are currently investigating other pulsation modes: nonadiabatic modes and also dark matter density driven modes.

In short, the first stars to form in the universe may be dark stars
powered by dark matter heating rather than by fusion.  Our work indicates that
they may become very large (up to $10^7 M_\odot$) and bright (up to $10^{10} L_\odot$),
thereby detectable in upcoming JWST observations.
They may provide seeds for the many supermassive black holes found in the universe.  
The observational possibilities of discovering dark matter by finding these stars with JWST data
is intriguing.  Further, once DS
are found, one can use them as a tool to study the properties of WIMPs.

\begin{figure}
\includegraphics[width=\textwidth]{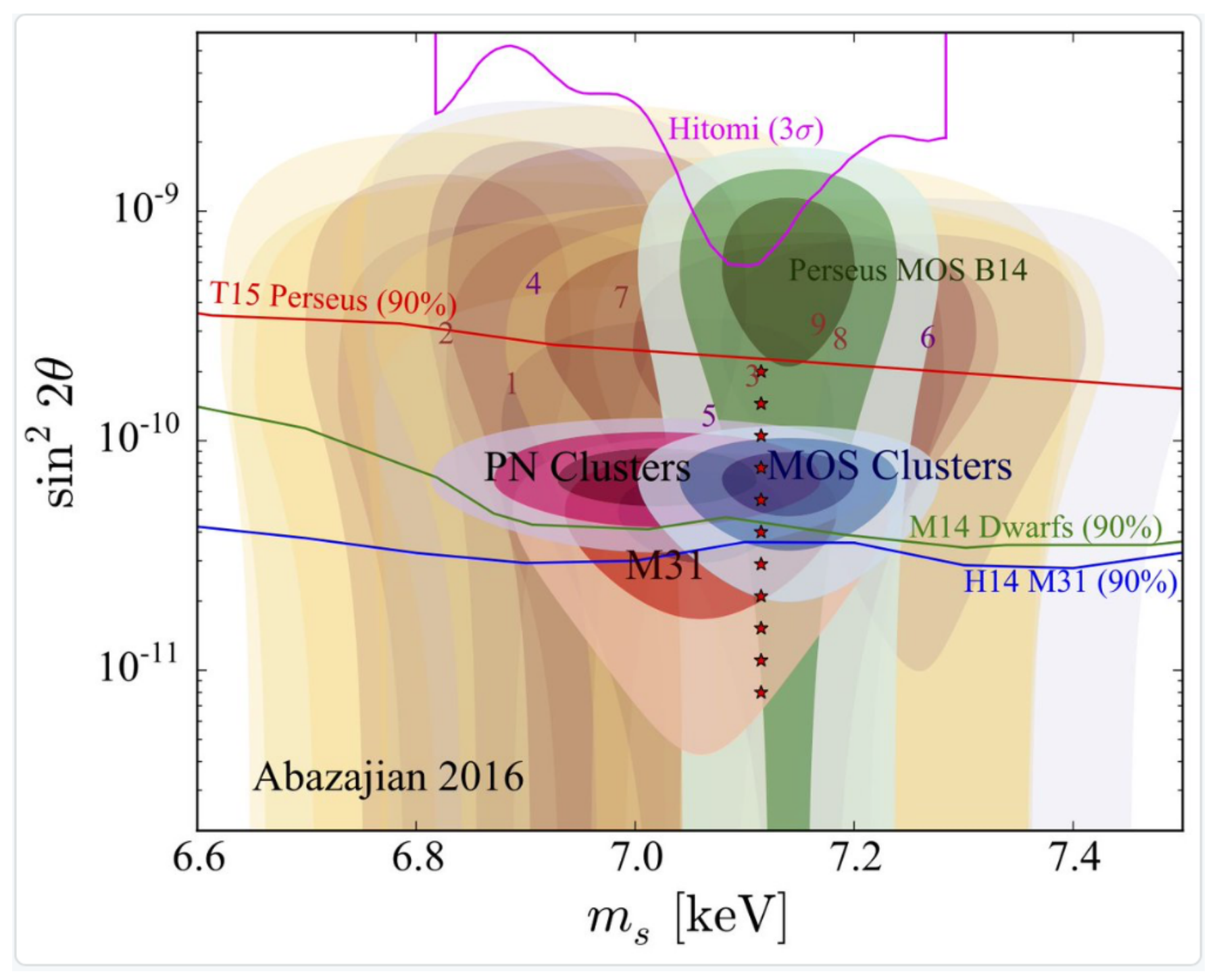}
\caption{ Sterile Neutrinos:
Observations consistent with and bounding the sterile neutrino mass and mixing angle to ordinary neutrinos
(figure courtesy of K. Abazajian ``Cosmology of Sterile Neutrinos," in preparation (2016)).}
\end{figure}

\section{Sterile Neutrinos}

Another intriguing dark matter candidate is a  sterile neutrino.
Whereas the three known neutrino species are far too light
to constitute dark matter, it is possible that one or more additional neutrino types, that do not
interact via the fundamental interactions of the standard model of particle physics
could make up the dark matter.
These sterile neutrinos could, however, mix with ordinary neutrinos.

In the past few years, several X-ray astronomy groups \cite{Bulbul:2014sua,Boyarsky:2014jta}
have found evidence for a 3.5 keV line in clusters
of galaxies and in M31.  This line would be consistent with a dark matter origin, corresponding to a 7 keV
rest mass sterile neutrino with vacuum mixing with active neutrinos
${\rm sin}^2 2 \theta \sim (2-20) \times 10^{-11}$.  Fig.~15 illustrates some of the observations.
However, others argue against this interpretation, e.g. Ref.~\refcite{Jeltema:2015mee} claims that the
line is not seen from the dwarf galaxy DRACO and thus the 7 keV sterile neutrino is ruled out.
This is a subject of
deep controversy.

Theoretical studies of sterile neutrinos are also ongoing.
The sterile neutrino is a singlet under the standard model; it is likely a right handed neutrino.
The production of these particles is difficult.  If thermal, they tend to overclose the universe.
Other mechanisms \cite{Barbieri:1990vx,Dodelson:1993je,Dolgov:2002wy}
 or resonance using a large lepton
asymmetry \cite{Canetti:2012vf}
are difficult but being investigated.
In many models the sterile neutrino constitutes warm dark matter, which leads to testable predictions such as the core/cusp
of galaxies and the numbers of substructures (objects smaller than our Galaxy), see e.g. Ref.~\refcite{Bozek:2015bdo}.

\section{What's Hot in Dark Matter}

As I've discussed, unexplained signals in a variety of data sets point to four hints of possible dark matter detection.
First,  the DAMA \cite{Bernabei:2014tea}
annual modulation \cite{dfs} signal could be compatible with a $\sim$10GeV WIMP.
However, since other experiments do not see any signal at all, the DAMA results must be checked.  Currently three
different experiments are planning to repeat the DAMA setup with NaI crystals:  SABRE, COSINE, and ANAIS.

Second, the Fermi-LAT $\gamma$-ray excess from the direction of the Galactic Center could be due to WIMP annihilation.
However, point sources (such as pulsars) constitute another explanation of the excess that is at least as good or better.

Third, the possible 3.5 keV X-ray line from clusters and from M31 could be explained by a 7 keV sterile neutrino, but this 
interpretation
is very controversial.

A fourth intriguing signal, not yet mentioned in this article, is the 511 keV $\gamma$-ray line in INTEGRAL data. 
\cite{Knodlseder:2003sv,Jean:2003ci}
This is seen in the Galactic Bulge out to 6 degrees (3 kpc).  There is no clear astrophysical explanation.
Low mass X-ray binaries were thought to be a compelling explanation which is now being ruled out.
The explanation for the line could be dark matter annihilation to e$^+$e$^-$ pairs.  This would be MeV dark matter.
\cite{Boehm:2003bt}

The future holds interesting studies of these signals as well as the continuing hunt for dark matter.

\section{Conclusion}
Most of the mass in the universe is in the form
of an unknown type of dark matter. The need for dark matter has become more
and more clear since the 1930s, with evidence from rotation curves,
gravitational lensing, hot gas in clusters, the Bullet Cluster, structure formation, and
the cosmic microwave background.    A consensus picture has emerged, in which the dark matter
contributes 26\% of the overall energy density of the universe.  Its
nature is still unknown.  At most 15\% of the dark matter in galaxies can be
white dwarfs (or other MACHO candidates), but most is likely to be an
exotic particle candidate.  Dark matter searches for the best motivated
candidates, axions and WMPs are ongoing and promising over the next
decade.

The interesting unexplained signals that may herald the discovery of dark matter have
been reviewed:  DAMA's annual modulation signal and the
Fermi-LAT gamma-rays from the Galactic Center might be due to WIMPs, a 3.5 keV X-ray line from various
astrophysical sources is possibly from sterile neutrinos,
and the 511 keV line in INTEGRAL  might be due to MeV dark matter.
All of these would require further confirmation in
other experiments or data sets to be proven correct.
In addition, a new line of research on dark stars was reviewed
which suggests that the first stars to exist in the universe
were powered by dark matter heating rather than by fusion:
the observational possibilities of discovering dark matter by finding these stars with JWST data
 were discussed.  The goal of the searches over the next decade is to decipher the nature of the unknown dark matter.

\section{Acknowledgments}
KF would like to thank Luca Visinelli for commenting on the draft.
KF acknowledges support through a grant from the Swedish Research Council (Contract No. 638-2013-8993). KF  acknowledges support from DoE grant DE-SC007859 at the University of Michigan.

\end{document}